\documentclass[journal]{IEEEtran}

\usepackage{graphicx}
\graphicspath{{./figures/}}
\usepackage{epstopdf}
\DeclareGraphicsExtensions{.eps}
\usepackage{amsmath}
\usepackage{color}
\usepackage{algorithm}
\usepackage{algorithmic}
\usepackage{tabularx}
\usepackage{tabu}
\usepackage{pbox}
\usepackage{multirow}
\usepackage{url}
\usepackage{amssymb}

\begin{document}
%
\title{IoTSign: Protecting Privacy and Authenticity of IoT using Discrete Cosine Based Steganography}

\author{Alsharif~Abuadbba,
        ~Ayman~Ibaida, and~Ibrahim~Khalil
\IEEEcompsocitemizethanks{\IEEEcompsocthanksitem Alsharif Abuadbba is with Data61 CSIRO Australia, Ayman Ibaida is with Victoria University Australia and Ibrahim Khalil is with the RMIT University Australia.\protect\\
E-mail: sharif.abuadbba@data61.csiro.au, Ayman.Ibaida@vu.edu.au and ibrahim.khalil@rmit.edu.au}
}


\IEEEcompsoctitleabstractindextext{%
\begin{abstract}
Remotely generated data by Intent of Things (IoT) has recently had a lot of attention for their huge benefits such as efficient monitoring and risk reduction. The transmitted streams usually consist of periodical streams (e.g. activities) and highly private information (e.g. IDs). Despite the obvious benefits, the concerns are the secrecy and the originality of the transferred data. Surprisingly, although these concerns have been well studied for static data, they have received only limited attention for streaming data. Therefore, this paper introduces a new steganographic mechanism that provides (1)  robust  privacy protection of secret information by concealing them arbitrarily in the transported readings employing a random key, and (2) permanent proof of originality for the normal streams. This model surpasses our previous works by employing  the Discrete Cosine Transform to expand the hiding capacity and reduce complexity. The resultant distortion has been accurately measured at all stages - the original, the stego, and the recovered forms - using a well-known measurement matrix called Percentage Residual Difference (PRD). After thorough experiments on three types of streams (i.e. chemical, environmental and smart homes), it has been proven that  the original streams have not been affected ($<$ 1 \%). Also, the mathematical analysis shows that the model has much lighter (i.e. linear) computational complexity $\mathcal{O} (n)$ compared to existing work.

\end{abstract}

}

\maketitle

\IEEEdisplaynotcompsoctitleabstractindextext

\IEEEpeerreviewmaketitle

\section{Introduction}
\begin{figure}[!h] 
  \centerline
  {\includegraphics[scale=0.3]{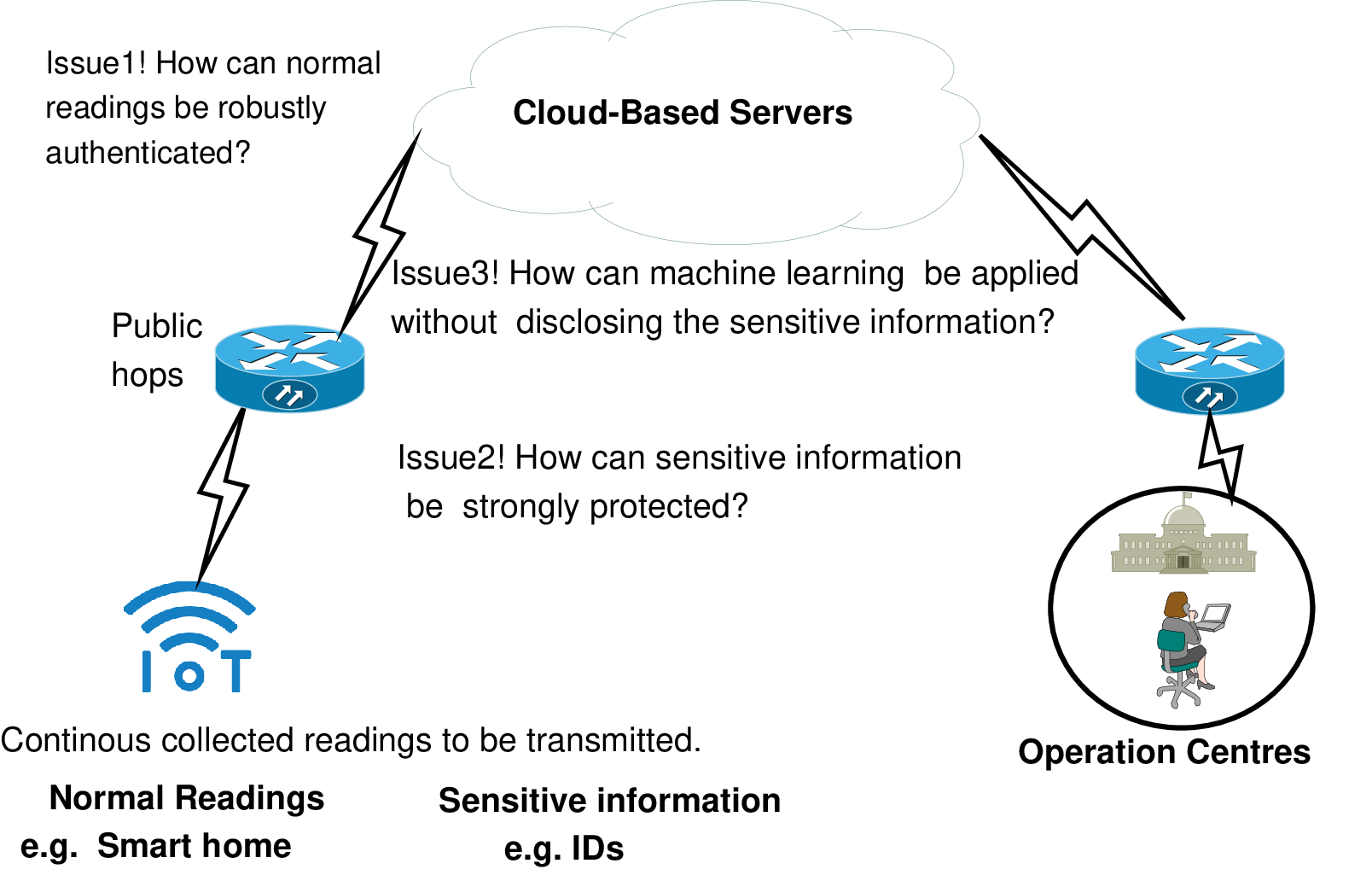}}
  \caption{Notable issues arise when periodically collected readings by remote IoT should be directly sent to cloud-based servers for analysis.} 
 \label{fig:ExistingIssues}
  \hfil
  \end{figure}

Lately, enormous interest has been expressed in gathering data remotely to effectively monitor various activities such as climate change, border invasion, battlefield scenarios, nuclear facilities or traffic screening \cite{int:yick2008wireless,ibaida2021privacy,baniata2021energy}. The streams are collected wirelessly using small sensors called Internet of Things (IoT) and forwarded to their final destination (e.g. operation centers). The continuous streams usually contain two types of data:  (1) normal samples (e.g. activity data) and (2) extremely secret information (e.g. nuclear facility or border screen geometric location, facility IDs and small pictures of the locations coupled with date and time). Despite the apparent  advantages of these activities, they create various concerns for the privacy of the secret information and the authenticity of the ordinary readings (See Fig. \ref{fig:ExistingIssues}). Surprisingly, while these concerns have been well studied for static data \cite{PrivToStatic:kargupta2003privacy,PrivToStatic:liu2006random,PrivToStatic:machanavajjhala2007diversity}, they have received limited attention for streaming data \cite{PrivStream:li2007hiding}.

The reasons for this are (1) the generators of these streams (i.e. remote IoT) pose unique challenges (e.g. their presence in an uncontrollable environment, resource constraints such as memory and power, and topological constraints where the data should go through multiple public hops to the final destination) which prevent a direct transplant of existing privacy protection and authenticity techniques \cite{sensorissue:li2009privacy}; (2) the massive  size of these streams  force the operation centers to do offshore operations (e.g. using cloud servers). 

To overcome the identified privacy and authenticity concerns, simple encryption-based techniques using symmetric and asymmetric keys and digital signature have been proposed \cite{tradcrypto:yang2008sdap,tradcrypto:lu2013lightweight,tradcrypto2013:Junzuo,tradcrypto:puttaswamy2014preserving,tradcrypto2014:Han,tradcrypto2017:song}. However, their main limitation is that machine learning techniques and mathematical operations cannot be applied directly to the encrypted form (i.e. cipher-text) which leads to revealing the keys to the intermediate hops and cloud providers to efficiently work on the data (i.e. the confidentially issue). 

The confidentiality issue has been overcome by applying a different type of cryptography called  homomorphism \cite{hom:liang2013udp,hom:lien2013novel,hom:saleem2014aggregation,hom:kumar2015secure,homocrypto2019:Bor}. The merit of homomorphic cryptography is that it allows certain mathematical operations to be applied to the cipher-text without exposing the keys. However, homomorphic encryption techniques still cannot work with many non-linear functions.  They are also still very complex and  not practical \cite{wat:zhang2008secure}.

Further, the widely used approaches with static data to maintain a reasonable level of privacy, which are based on random perturbation \cite{randper:kargupta2003privacy,randper:huang2005deriving} and $K$-anonymity \cite{k-anonymity:gedik2005location,homcrisis:groat2011kipda,k-anonymity:chow2011privacy} cannot solve both identified issues together: i.e. the privacy of the secret information and the authenticity of the ordinary readings. In fact, their concerns are slightly different where the random perturbation is efficient in a scenario of a central data collector gathers and mines  data from multiple providers, whereas  $K$- anonymity is mainly used to obfuscate published data \cite{sensorissue:li2009privacy}. 

Therefore, researchers are obliged to look for new solutions to do the following tasks. (1) Provide solid  privacy protection for the remotely collected secret data (e.g. geometric location, location picture, IDs and time). (2) Provide permanent evidence of authenticity for the non-sensitive collected readings. (3) Allow machine learning techniques and mathematical operations to be easily applied without exposing the secret information.

Another candidate technique to protect private information and seal transported data is steganography where a secret message is concealed within a larger transferred content (e.g. image or signal) in such a way that only the intended recipient can recover it \cite{stego:provos2003hide}.  However, simple steganography alone has two limitations. (i) It cannot ensure controlled access to the secret information because the hiding is performed in fixed positions (i.e. confidentiality issue).  (ii) It cannot conceal large amounts of secret information (i.e. capacity issue). Although steganography has been widely studied and used in the multimedia domain \cite{stego:cox2002digital}, we demonstrated how it can be utilized to protect the privacy and authenticity in the context of streaming  data \cite{stego:abuadbba2015wavelet,stego:abuadbba2015robust,stego:abuadbba2016resilient}. However, due to the employed signal processing types (e.g. Walsh-Hadamard and wavelet), (1) the size of hidden information was very limited and (2) computational complexity was very high (e.g. $\mathcal{O} (n^2)$) rendering them unsuitable in a limited resource  environment.

\subsection{Contribution}
\begin{itemize}
    \item We introduce a novel steganographic based privacy protection model surpasses existing models in terms of the appropriate balance between the volume of embedded secret data (i.e. up to 10 bits per coefficients), which was about 5-6 bits in our previous works \cite{stego:abuadbba2015robust,stego:abuadbba2016resilient}, and the resultant distortion (i.e. $<1\%$).  This is due to the efficient utilization of DCT signal processing.  This has been examined in details and showed in Section \ref{sec:experiments}.
    \item We improve the mathematical computational complexity where we achieved a linear steganographic algorithm $\mathcal{O} (n)$ by employing a simple 1D DCT technique while improving the hiding security  into two dimensional space.  This has been mathematically investigated and presented in Section 5.
\end{itemize}

\begin{table*}
  \centering
  \caption{Related models Summary.}
  \label{tb:relatedworksummary}
 \begin{tabu}{ |X[1.5]|X[4]|X[2.5]|} \hline
    Protection mechanism & Characteristics & Notes \\ \hline
    Simple Cryptography  \cite{tradcrypto:yang2008sdap,tradcrypto:lu2013lightweight,tradcrypto2013:Junzuo,tradcrypto:puttaswamy2014preserving,tradcrypto2014:Han,tradcrypto2017:song}. & 
   \begin{itemize}
             \item {}Apply symmetric/asymmetric encryption at  the remote side.
             \item {}All data should be decrypted to perform mathematical operations. 
             \item {}All data have to be decrypted before usage.
   \end{itemize}  &  
   \begin{itemize}
  \item {} Weak Confidentiality. 
  \item {} Low Efficiency. 
  \item {} Keys' management trouble.
  \end{itemize} \\ \hline 
    Homomorphic Cryptography \cite{hom:liang2013udp,hom:lien2013novel,hom:saleem2014aggregation,hom:kumar2015secure,homocrypto2019:Bor}. & \begin{itemize}
               \item {}Apply homomorphic crypto at the remote side.
               \item {} The obfuscated form of data can be utilized.  
               \item {}All data have to be decrypted before usage. 
     \end{itemize}  & 
   \begin{itemize}
    \item {}Strong Confidentiality.
    \item {}Not working with non-linear functions.
    \item {}Not applicable in real applications.
    \item {}Low Efficiency.
    
    \end{itemize} \\ \hline
    Random Perturbation \cite{randper:kargupta2003privacy,randper:huang2005deriving}. &  
   \begin{itemize}
                 \item {Distribute noise along with the principal components.}               
                 \item {Intermediate hops and Clouds can Not work accurately on the streams.}  
                 \item {Legitimate recipient should take out the noise to fully understand.} 
       \end{itemize} & 
   \begin{itemize}
   \item {} Not optimal with numeric and non-stationary streams \cite{PrivStream:li2007hiding}.
   \item {}No Confidentiality.
   \item {}Weak Authenticity.
   \item {}Medium Efficiency.

      \end{itemize} \\ \hline  
  $K$- Anonymity \cite{k-anonymity:gedik2005location,k-anonymity:chow2011privacy}. &  
    \begin{itemize}
                  \item {Obfuscate the remote sender within a larger group. }               
                  \item {Intermediate hops and clouds can not work accurately on the readings.}  
                  \item {Focus on publishing an obfuscated-approximate data's version for public usage.} 
        \end{itemize} & 
    \begin{itemize}
    \item {} Has subtle privacy issues \cite{k-anonymityCris:machanavajjhala2007diversity}.
    \item {}Weak Confidentiality.
    \item {}Weak Authenticity.
    \item {}Low Efficiency.

       \end{itemize} \\ \hline 
     Steganography \cite{stego:abuadbba2015wavelet,stego:abuadbba2015robust,stego:abuadbba2016resilient}. &  
    \begin{itemize}
    	\item {Embed secret information inside IoT streams. }               
    	\item {Intermediate hops and clouds can work accurately on the stego readings.}  
    	\item {Using frequency domain for hiding.} 
    \end{itemize} & 
    \begin{itemize}
    	\item {} Low hiding capacity.
    	\item {} Non-linear computational complexity.    	
    \end{itemize} \\ \hline
 \end{tabu}
  \end{table*}
In the proposed model, the end-point IoT private information is  randomly concealed inside the periodically collected normal streams and only authorized parties can recover this private data. This is achieved as follows. DCT is applied to the ordinary stream. The resultant coefficients are then reshaped to $2D$ $M \times N$ matrix employing an integrated random  key (at the remote IoT side). Then, the key is applied to (1) only obfuscate the secret data (e.g. IDs, geometric location and location picture coupled with date and time) using a fast symmetric encryption, and (2) create an arbitrary sequence of coefficients which will be followed to conceal the private information. After the hiding process, an inverse DCT is employed to recompose the normal stream (i.e. stego readings) and transport them. Hence, only the parties who have the key (e.g. operation centers) can recover and decrypt the private data. The other benefit is that, the stego form of IoT stream can be utilized without removing the secret bits. Therefore, our algorithm will allow machine learning techniques and mathematical operations to be easily applied on stego streams without exposing the secret information.

\section{Related Work} 
Today, any proposed solution to the remotely collected IoT streams should carefully contemplate two main characteristics: security (i.e. solid end-to-end protection of the transmitted secret information and the authenticity of the transported streams) and efficiency (i.e. allowing mathematical operations to be directly applied without confidential information disclosure). However, these features are poorly balanced in most current solutions. Table \ref{tb:relatedworksummary} recaps most of the relevant work and classifies them into five categories based on the applied mechanisms: using simple, homomorphic cryptography, random perturbation and K-anonymity.

\section{Methodology}
A balanced level between the two main concerns (i.e. security and efficiency) has been carefully considered in the introduced steganographic algorithm as follows: (1) it is highly improbable for unauthorized parties to discover the concealed data without a relevant key, and (2) there is little distortion on the transmitted streams, so they can be directly used without removing the hidden private data.

The operations at the deployed IoT can be classified into three stages.

\subsection{The Discrete Cosine Transform}\label{subsec:dcttransform}
DCT is a widely-known transformation technique that can convert a stream into a set of frequency components called coefficients \cite{dct:feig1992fast,dct:rao1990discrete}. This is important as it allows the resultant values to be categorized into an aware energy distribution where most of the important values can approximate the signal. The significant feature of this mechanism is that the actual signal can nearly be recomposed from only a few coefficients.

\begin{figure*}[!t]
 \centerline
 {\includegraphics[scale=0.4]{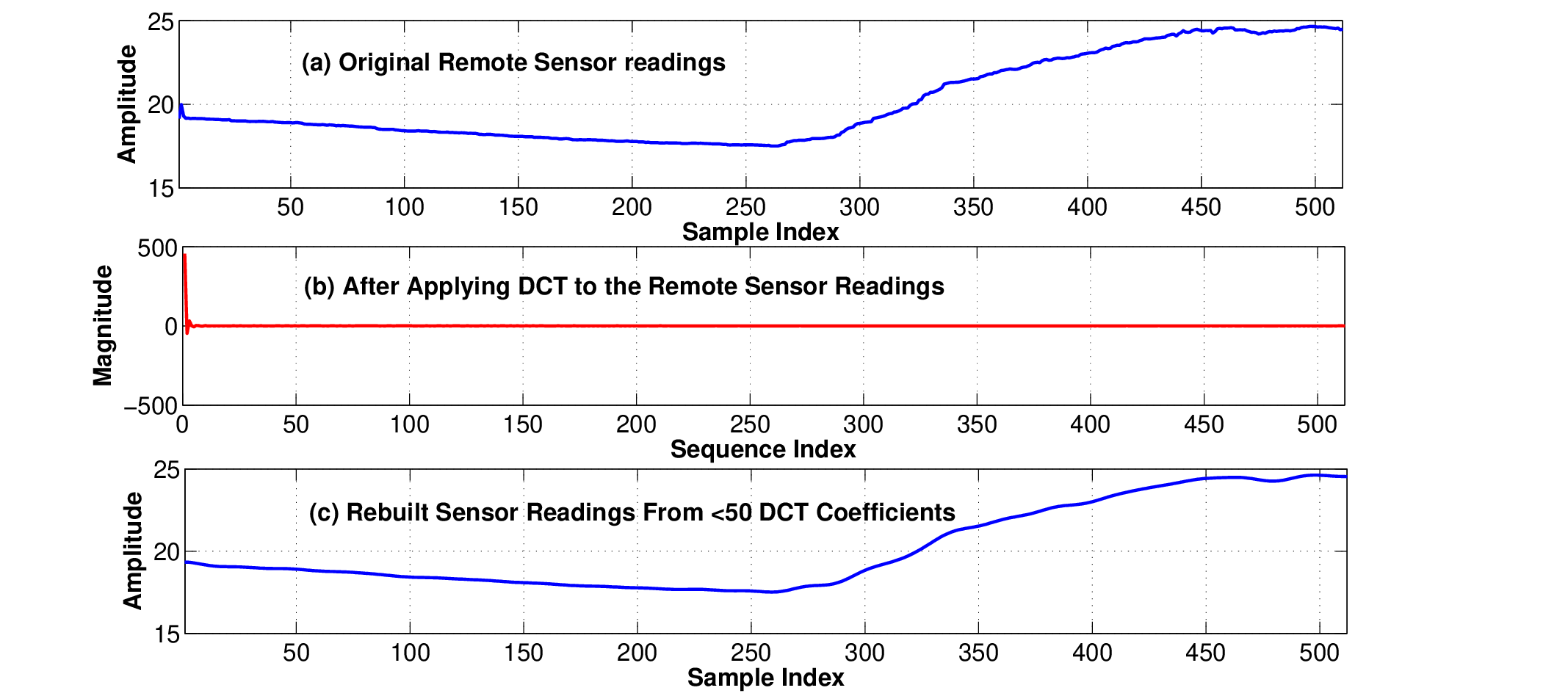}}
 \caption{IoT readings: (a) Direct plot, (b) After applying DCT and (c) recomposed form after zeros more than 95 \% of DCT coefficients.}
 \label{fig:DCTeffecincy50coef}
 \end{figure*}

For a visual demonstration, Fig \ref{fig:DCTeffecincy50coef} illustrates the impact of using DCT on a remotely collected stream. (a) The plot for $>$512 weather samples. (b) The resultant values after performing DCT, which obviously shows the distribution of the signal major energy (i.e. from 0 to $<$30), but others have lower importance. We accordingly wiped all coefficients between 30 and 512 to proof their impact on the recomposed stream. (c) The plot displays the recomposed original weather stream from only $<$30 coefficients. This obviously demonstrates the capacity and the flexibility that can be obtained from the less important coefficients. This influenced us to exploit DCT to conceal more private data related to remotely-distributed IoT without augmentation in the transmitted stream.
 
 \begin{figure*}[!h]
\centerline
{\includegraphics[scale=0.4]{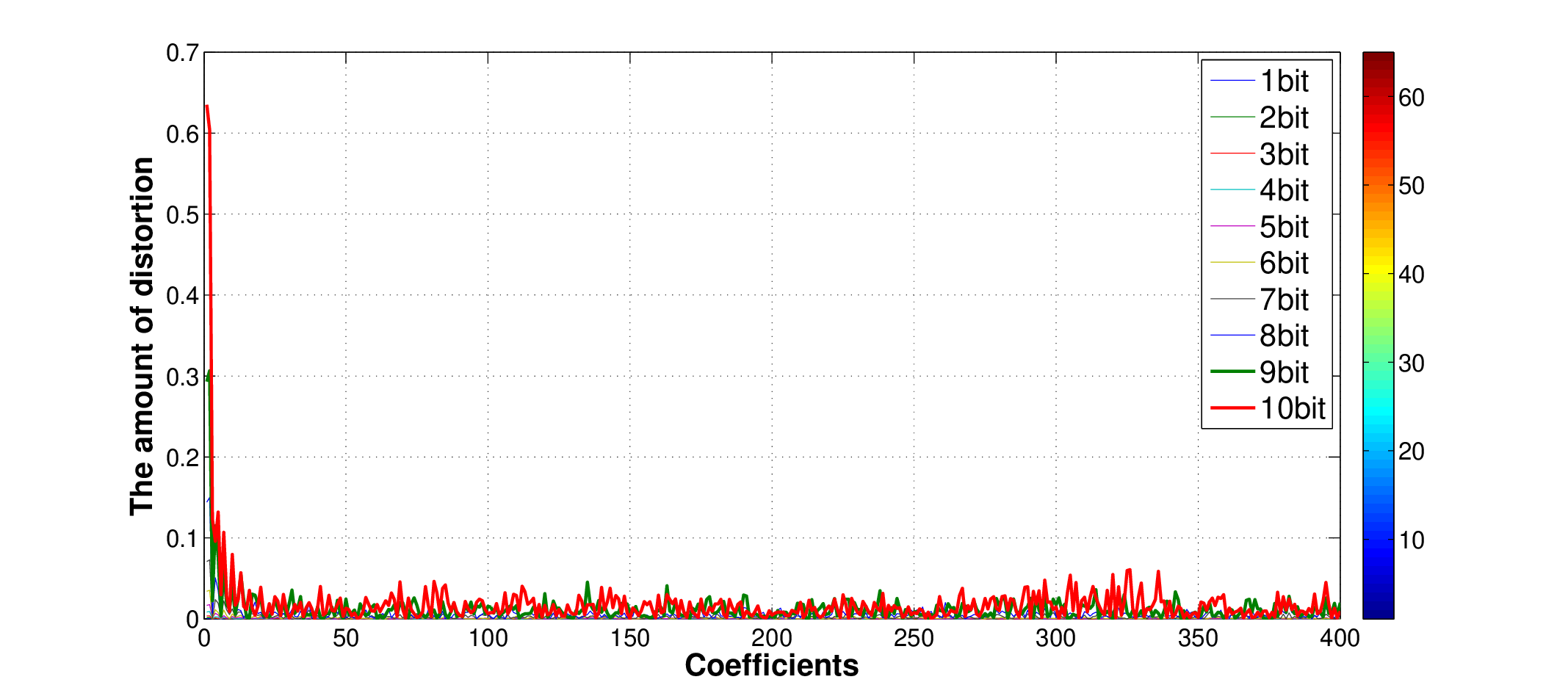}}
\caption{Resultant distortion after concealing different number (i.e. 1 to 10) of bits in each DCT coefficient.}
\label{fig:DCTamountofdistorion}
\hfil
\end{figure*}
 
There are several variations of DCT, but the most commonly used follows Ahmed et. al. \cite{dct:ahmed1974discrete}. It operates on a real sequence $x_n$ of length $N$ which results in an equal size sequence $y_k$ in a form of $C_k$ coefficients (See Eq \ref{eq:dct}). 

 \begin{equation}\label{eq:dct}
y(k)= w(k)\sum_{n=1}^{N}x(n) cos\left(\frac{\Pi (2n-1)(k-1)}{2N}\right),
 \end{equation}

where $k=1,...,N$ and,
 \begin{equation}\label{eq:dct1}
w(k)=\begin{cases}
\frac{1}{\sqrt{N}} & \text{ if } k= 1,\\ 
\frac{2}{\sqrt{N}} & \text{ if } 2\leq k\leq N. 
\end{cases}
 \end{equation}

DCT is selected for two main purposes. (1) Erasing or manipulating much of the resultant DCT coefficients will not impact the accuracy of the recomposed signal. (2) All resultant DCT's coefficients are real numbers which allows us to directly manipulate and reconstruct, maintaining a minimum distortion, whereas the resultant values from other signal processing mechanisms (e.g. Fast Fourier and Chirp Z) contain an imaginary part that hardens the embedding process and maximizes the distortion. (3) Unlike other transformation techniques (Walsh-Hadamard and Wavelet), DCT is very flexible where it can be applied to any stream length.

Consequently, in this model, DCT is performed on various real-time IoT streams remotely collected  from three different datasets - explained in Section \ref{sec:experiments} - which contains various readings such as chemical substances (e.g. ethanol), environment (e.g. temperature) and  smart home activities (e.g. power consumption). The output is then reorganized to a $2D$ matrix. The most significant values will not be altered due to their significant representation to the original readings. In contrary, certain bits will be manipulated in the rest of DCT coefficients. However, to ensure the lowest noise to the original  streams, many experiments have been done to choose a suitable steganography level (i.e. number of hidden bits  in each coefficient) as presented in Fig \ref{fig:DCTamountofdistorion}. The experiments prove that, up to $\leq$ 10 (ten) bits can be arbitrarily concealed in the less significant coefficients without noticeable distortion effect.

\subsection{Hiding}\label{subsec:hiding}
The private data will be concealed inside the resultant coefficients, after performing DCT to the collected streams. However, to ensure resilient security and to prohibit intruders from recovering this information, a security key is produced for every distributed IoT node and will be known only to the end receiver of the data. This key is used to enforce the following security layers.
\subsubsection{Confidential Information Encryption}\label{subsec:key}
\begin{figure}[!h] 
   \centerline
   {\includegraphics[scale=0.35]{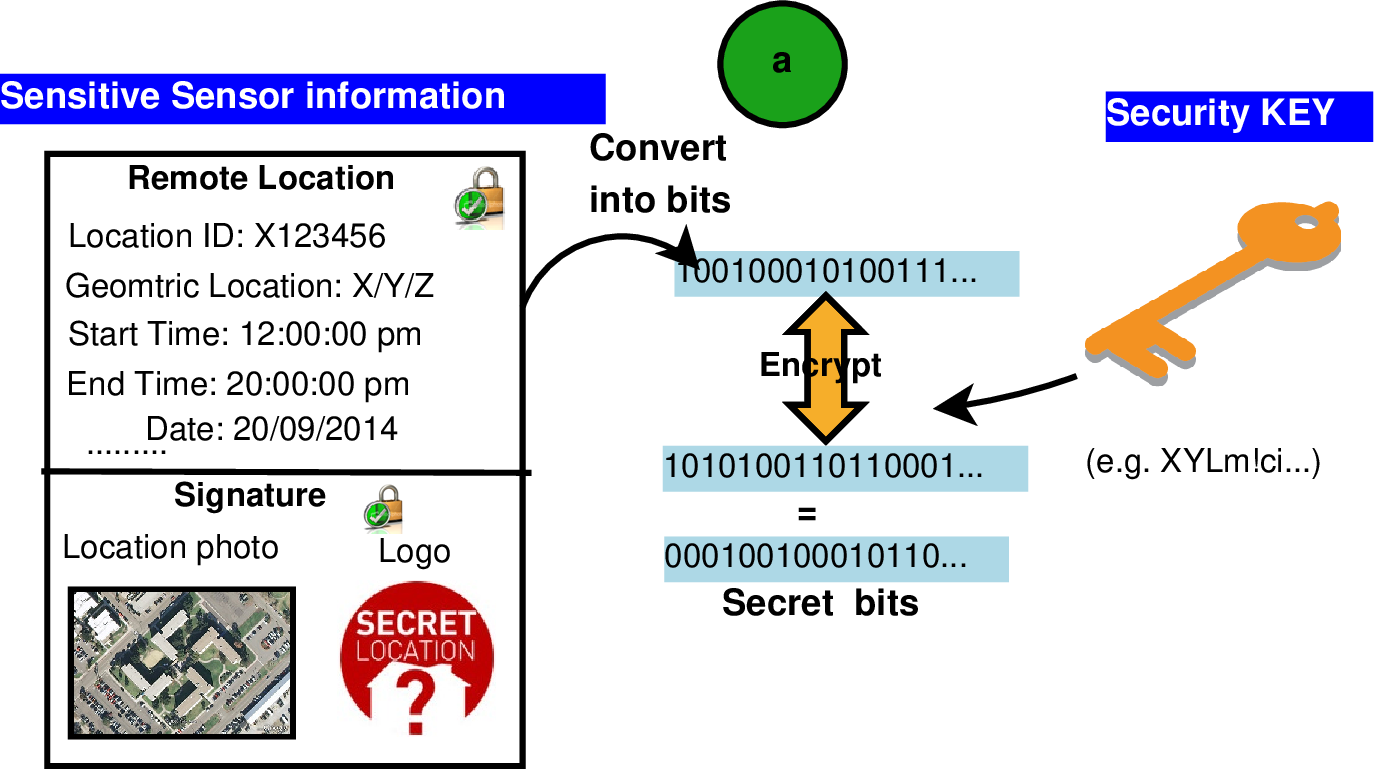}}
   \caption{An example of how the private information is obfuscated before hiding.}
   \label{fig:Aencryption}
   \hfil
   \end{figure}

The key is used to obfuscate the private information (e.g. IDs) before the hiding process using symmetric cryptography (e.g. AES), which is secure and suitable for  IoT's abilities (See Fig \ref{fig:Aencryption}). This is shown in Eq \ref{eq:xor}.
\begin{equation} \label{eq:xor}
\widetilde{C}\Leftarrow E(K,C)
\end{equation}
where $E$ is an AES algorithm, $K$ is the key, $C$ is the original private data and  $\widetilde{C}$ is the encrypted form.
\subsubsection{Coefficients Scrambling}\label{subsec:coefscramble}
\begin{figure}[!h] 
      \centerline
      {\includegraphics[scale=0.35]{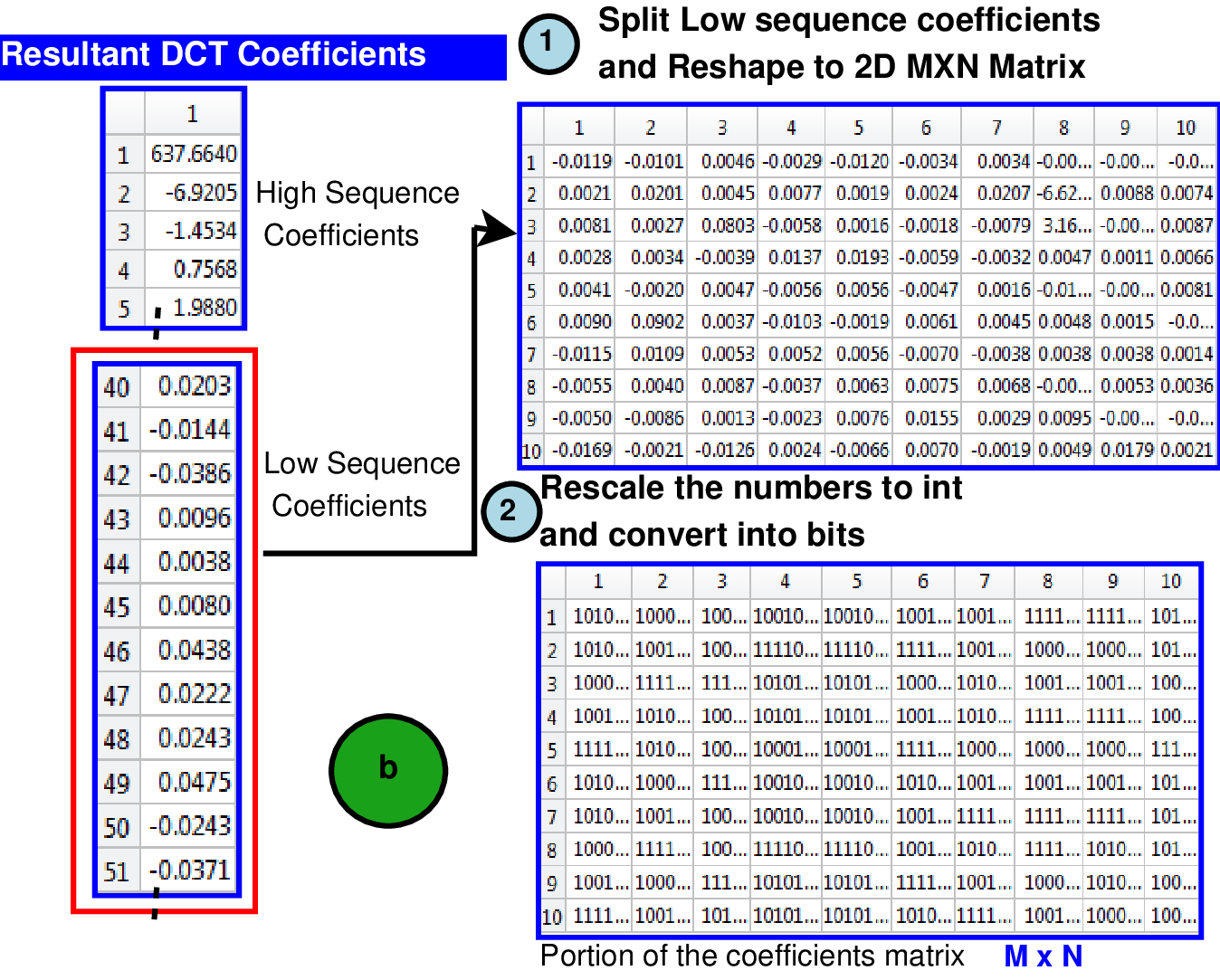}}
      \caption{Block diagram presents how the DCT coefficients are split, rescaled and converted into bits.}
      \label{fig:DCTreshaping}
      \hfil
\end{figure}

 The key is employed to scatter and reorganize the resultant DCT coefficients from a vector to $2D$  $M \times N$ matrix (See Fig \ref{fig:DCTreshaping}).
\subsubsection{Random Hiding Order}\label{subsec:randomorder}
The key is employed  to create a random sequence of coefficients in a form of $2D$ matrix that will be used to conceal the private information.  This is shown in Eq \ref{eq:selctedcoef}.

\begin{equation} \label{eq:selctedcoef}
\widetilde{N}\times\widetilde{M}\Leftarrow  f_x(K)
\end{equation}
where $\widetilde{N}\times\widetilde{M}$ is the created $2D$ sequence of coefficients and $f_x$ is the generation function.\\

\begin{figure}[!h] 
   \centerline
   {\includegraphics[scale=0.3]{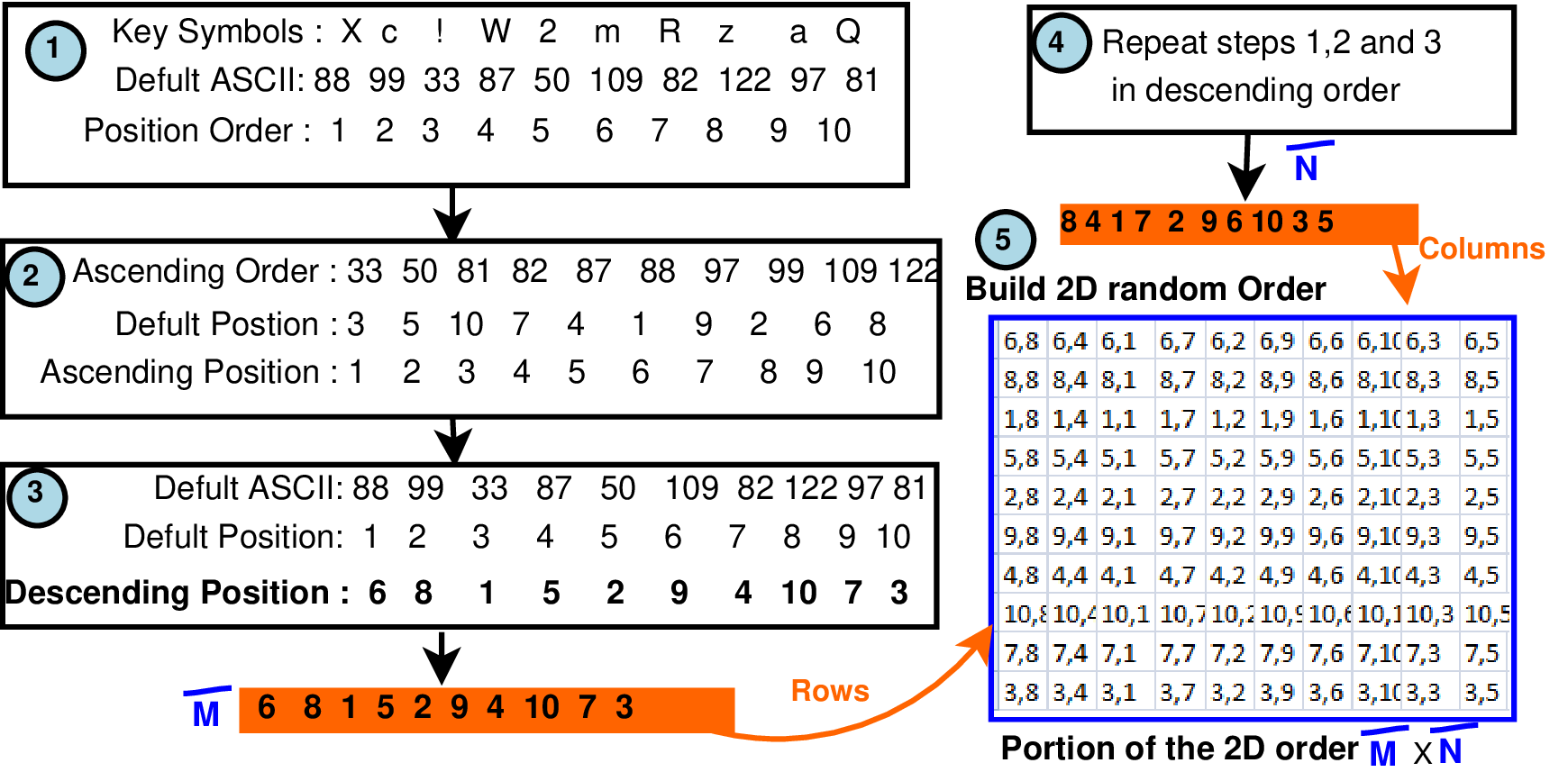}}
   \caption{Demonstration of a $2D$ concealing matrix order $\widetilde M \times \widetilde N$ that is created from the key.}
   \label{fig:genCoef}
   \hfil
   \end{figure}

For better understanding, Fig \ref{fig:genCoef} shows a demonstration of using the key to produce an arbitrary hiding sequence. Step 1: the key is transformed into ASCII i.e., Default ASCII and an initial order is distributed i.e., Position Order. Step 2: the ASCII  is then  ordered in ascending manner i.e., Ascending Order. simultaneously, shuffle the Default Position to keep track of the original ASCII locations. Then,   another sequence is assigned i.e., Ascending Position. Step 3: return the ASCII to its initial format by using the Default Position. A new unique sequence is ontained i.e.,  Descending Position $\widetilde M$. The reason of the 3 steps is to prevent two identical sequences from various keys.  Step 4: by reversing the order, almost identical steps i.e., 1-3 are reiterated to generate $\widetilde N$. Finally, $\widetilde M$ and  $\widetilde N$  are combined to compose the $2D$ matrix sequence. However,  the key is much longer (e.g.  $\geq 128$) in the proposed model.\\

These three steps will guarantee that only an authorized receiver who has the security key can recover and decrypt the private information properly.

\begin{figure}[!h] 
 \centerline
   {\includegraphics[scale=0.4]{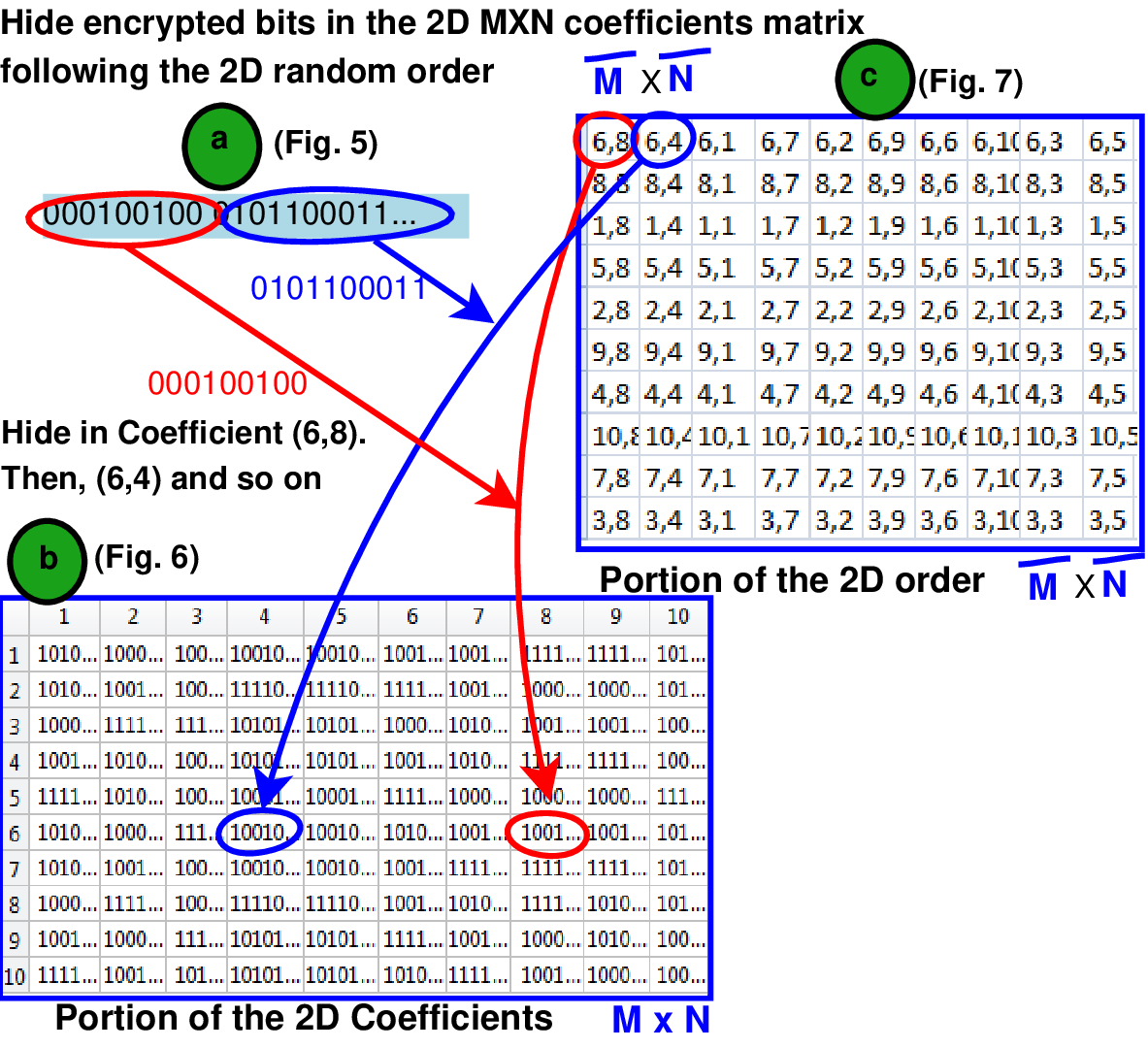}}
   \caption{Block diagram summarizes the hiding process and use the information explained in Figures \ref{fig:Aencryption}, \ref{fig:DCTreshaping} and \ref{fig:genCoef}.}
    \label{fig:Embeddingsteps}
     \hfil
\end{figure}

The detailed process of hiding is shown in Fig \ref{fig:Aencryption}, \ref{fig:DCTreshaping}, \ref{fig:genCoef} and summarized in Fig \ref{fig:Embeddingsteps}. After applying DCT to the ordinary stream, the output is scattered and recomposed to $M \times N$ $2D$ matrix. Then, the key is  employed to obfuscate the private information. After that, the key is harnessed to create the random $2D$  sequence. The private bits will be then concealed corresponding to this order. The summary of all steps in this process is shown in section \ref{subsubsec:hidesummary}.

\subsection{Hiding Steps Summary}\label{subsubsec:hidesummary}
\begin{enumerate}
\item The private information is classified and obfuscated (i.e. using the key) $\Rightarrow$ secret bits.
\item Hiding order is created (i.e. using the key)  $\Rightarrow$ 2D $\widetilde M \times \widetilde N$.
\item DCT  is applied on normal streams.
\item Least important coefficients are selected. Based on our observations from all experiments on three different datasets, the important coefficients were $\le 20\%$  because of the nature of the data in this context. 
\item Rescale these coefficients $\Rightarrow$ integer why?

To make sure the selected coefficients are ready for steganography, they should be in an integer format and to avoid the issue of differentiating between negative and positive, all numbers should be positive. This may be done by adding a threshold $\varphi$ and multiplied by $\vartheta$ to maintain all their details (i.e. four decimal positions).
\item DCT coefficients  are scattered $\Rightarrow$ 2D $ M \times  N$.
\item  The hiding process is started. 
\item After finishing, the resultant coefficients recomposed then rescaled  by dividing all stego coefficients by $\vartheta$ and subtracting the threshold $\varphi$. 
\item Re-combine coefficients and apply invrse DCT.

\end{enumerate} 
\subsection{Inverse Discrete Cosine Re-Transform}\label{subsec:idct}
The output of the embedding stage is named stego coefficients.  The stego coefficients at this point is re-scattered to a vector, and the Inverse DCT applied to re-transform the collected streams to the initial time domain. The output of this process is Stego stream (i.e. comprises concealed private data) which is almost similar to the original stream. The advantage of that is - as will be shown in Section \ref{sec:experiments} - the stego streams can be immediately utilized. However, only the key holder parties  can recover the hidden data  and validate them. The DCT inverse can be formulated as in Eq \ref{eq:idct}.

 \begin{equation}\label{eq:idct}
x(n)= \sum_{n=1}^{N}w(k)y(k) cos\left(\frac{\Pi (2n-1)(k-1)}{2N}\right)
 \end{equation}
 
where $k=1,...,N$ and, 
 \begin{equation}\label{eq:idct1}
w(k)=\begin{cases}
\frac{1}{\sqrt{N}} & \text{ if } k= 1,\\ 
\frac{2}{\sqrt{N}} & \text{ if } 2\leq k\leq N. 
\end{cases}
 \end{equation}
 
\subsection{Private Information Recovery}\label{subsec:retrieve}
\begin{figure}[!h] 
   \centerline
   {\includegraphics[scale=0.35]{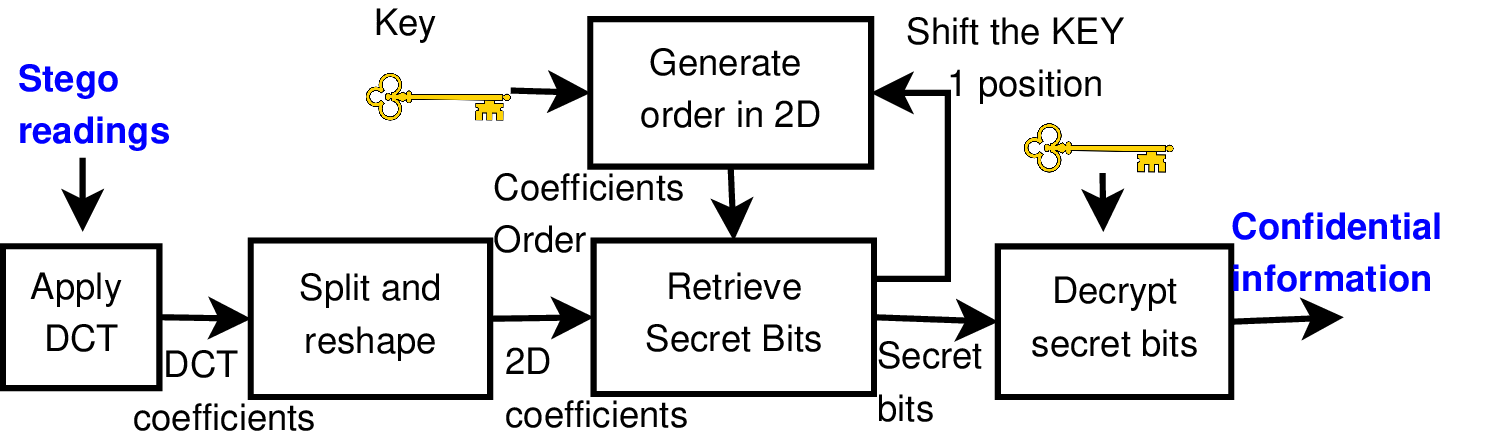}}
   \caption{The main steps to extract the sensitive information.}
   \label{fig:exractsteps}
   \hfil
   \end{figure}

To accurately extract and decrypt the private hidden bits, the security key has to be obtained. The process is nearly similar to the concealing steps, but the bits will be recovered rather than  embedded. Fig. \ref{fig:exractsteps} demonstrates the recapped steps. First, DCT is applied to stego  stream. The key is then employed to reorganize the DCT coefficients into a $2D$ form and create the selected coefficients' sequence. Next, the secret bits' recovery is started, corresponding to the produced order. Finally, the key will  be used  to decrypt the secret bits and verify the resultant information.
\section{Evaluation} \label{sec:evaluation}
This section concentrates on the proposed model assessment from various angles such as the key robustness, hidden data security, the supreme size of private information that can be concealed and the distortion measurements.  
\subsection{Key Robustness}
Both transmitted streams and the private key is required to be revealed in advance to recover the concealed secrets by legitimate parties. On the other hand, an anonymous intruder has to be aware of the existence of hidden message in addition to performing a brute force attack to break the model, which makes it extremely difficult.  

However, the security key plays an important role in our model due to its usage to ensure various security layers (See Figs. \ref{fig:Aencryption}, \ref{fig:DCTreshaping} and \ref{fig:genCoef}): (1) obfuscate the private information (2) reorganize DCT values into $2D$ $M \times N$ matrix,  and (3) create a random coefficients' sequence as $2D$ $\widetilde M \times \widetilde N$ matrix  to conceal the bits. Therefore, the secrecy of this key is extremely important. The two involved parties should maintain this key very carefully.  (a) At the first party (i.e. remote IoT), the key has to be integrated, and (b) the second party (e.g. operation centers) has to protect this key and employ it to recover and check the  validity of the hidden private information whenever is required. Accordingly, Only stego streams are visible to other parties.

The key strength of our model can be quantified as the  entropy bits' number $H$ (See Eq. \ref{eq:keylength}) where $2^H$ is the supreme possibilities that would need to be examined by anomalous intruders during a brute-force attack.
\begin{equation}\label{eq:keylength}
H=log_2N^L
\end{equation} 
Where $L$ is the total symbols length and $N$ is symbols' probabilities. Table II presents a demonstration  of  different key lengths, key symbols sets and the total number of their possibilities.

\begin{table}[!h]  \label{tb:keystrengths}
   \caption{ Example of Different used keys strength}
\begin{center}
  \begin{tabular}{| c | c | c | }
    \hline
    Key length & Symbol Set & Possibilities\\ \hline
    64 & US-ASCII & 7.2e+134 \\ \hline
    128 & US-ASCII &  5.2e+269 \\\hline
    256 & US-ASCII &  $\infty$\\\hline
    64 & UTF-8 & 1.3e+154 \\ \hline  
    128 & UTF-8 & $\infty$ \\\hline
    256 & UTF-8 & $\infty$ \\\hline
    64 & UTF-16 & $\infty$ \\ \hline       
    128 & UTF-16 & $\infty$ \\\hline
    256 & UTF-16 & $\infty$ \\\hline
  \end{tabular}
\end{center} 
\end{table}
\subsection{Unauthorized Retrieval}
To protect the private embedded bits  from brute-force recovery, the reorganized $2D$ $M \times N$ coefficients after DCT transformation of the collected streams have to be higher than arbitrary size (e.g. $>$ Key volume) (See Eq \ref{eq:nofcombination}). 
\begin{equation}\label{eq:nofcombination}
T_p=\sum_{i=t}^{r}R! \times \sum_{j=t}^{c}C! \times N^L
\end{equation}
Where $T_p$ is the combinations entropy, $C$ and $R$ are the reorganized $2D$ coefficients  and  $t$ is an offset which highlights the lowest entropy of each row selection. $r$ refers to the highest entropy of the rows in $R$$\times$$C$ matrix, and identically $c$ refers to the maximum number of columns in that matrix.

For instance, assume collected streams of size 512 points, and reorganized to $2D$ coefficients  of  $32 \times 16$ after performing DCT. The chosen offset is $16 \times 8$, the key charter set is UTF-8 and its volume is 128 (See Eq \ref{eq:nofcombinationexample}).

\begin{equation}\label{eq:nofcombinationexample}
T_p=\sum_{i=1}^{32}32! \times \sum_{j=16}^{32}32! \times 256^{128} \Rightarrow T_p=\infty
\end{equation}

Consequently, this confirms that recovering and decrypting the intended private data  properly in a reasonable time is highly improbable.
\subsection{Hidden Data Size}

The maximum amount of hidden data in $X$-stream mainly depends on two values: (1) the stream's size $X$ (e.g. temperature) and (2) the hidden bits per coefficient. However, if we use the streams directly even by hiding 1 bit per reading, the distortion will be high. One of the possible solutions is using signal processing to transform the readings from their time domain to frequency domain. Based on that, another dimension becomes very important to the hiding capacity which is the coefficients' selection process.  Therefore, this feature has been exploited in our model to exclude few important coefficients and hide more data in others to maximize the capacity with maintaining low distortion. In the proposed algorithm, the maximum number of bits that can be hidden in $X$ transmitted stream with maintaining the lowest distortion impact is shown in Eq \ref{eq:sizeofhiddendata}. 
\begin{equation}\label{eq:sizeofhiddendata}
b=\sum_{i=1}^{n} ((R \times C)-h )\times B
\end{equation}
Where $b$ is the highest entropy of hidden bits, $n$ is the total streams samples, $R$ and $C$  are the rows and columns of the reorganized $2D$ coefficients after applying DCT  transformation to the original values, $h$ is the number of high sequence coefficients and $B$ is the concealed bits in each coefficient.

For instance, suppose that DCT transformation is exercised to ordinary stream, and  the size of the reorganized $2D$ matrix is $16 \times 512$ (i.e. $R$ and $C$). Also, suppose the high sequence coefficients is $\leq $50 (i.e. the value of $h$)  and about 9 bits (i.e. value of $B$) are concealed in each coefficient. Consequently, nearly  9159 bytes (9 KB) of private data can be concealed within these coefficients.

\subsection{Stego Efficiency }
To precisely evaluate our model's impact on the  collected streams, the margin between the ordinary and stego streams  (i.e. resultant distortion)  has been thoroughly monitored using  percent of Percentage Residual difference (PRD). The PRD is a widely-used measurement that is known for its precision of detecting any recomposition error between the ordinary and the recomposed streams as defined in Eq. \ref{eq:prds} \cite{prd:cetin2000compression}.
\begin{equation}\label{eq:prds} 
PRD=\sqrt{\frac{\sum_{n=1}^{N}(x(n)-\widetilde{x}(n))^2}{\sum_{n=1}^{N}(x^2(n))}}\times 100
\end{equation}
where $x(n)$ and $\widetilde{x}(n)$ are the ordinary and the recomposed streams,  and $N$ is the stream's size.

The PRD measurement is also employed to accurately calculate the resultant distortion caused by the recovery process (i.e. between the ordinary and the extracted streams). All results are highlighted in Section \ref{sec:experiments}.

\begin{figure*}[!t]
\centerline
{\includegraphics[scale=0.4]{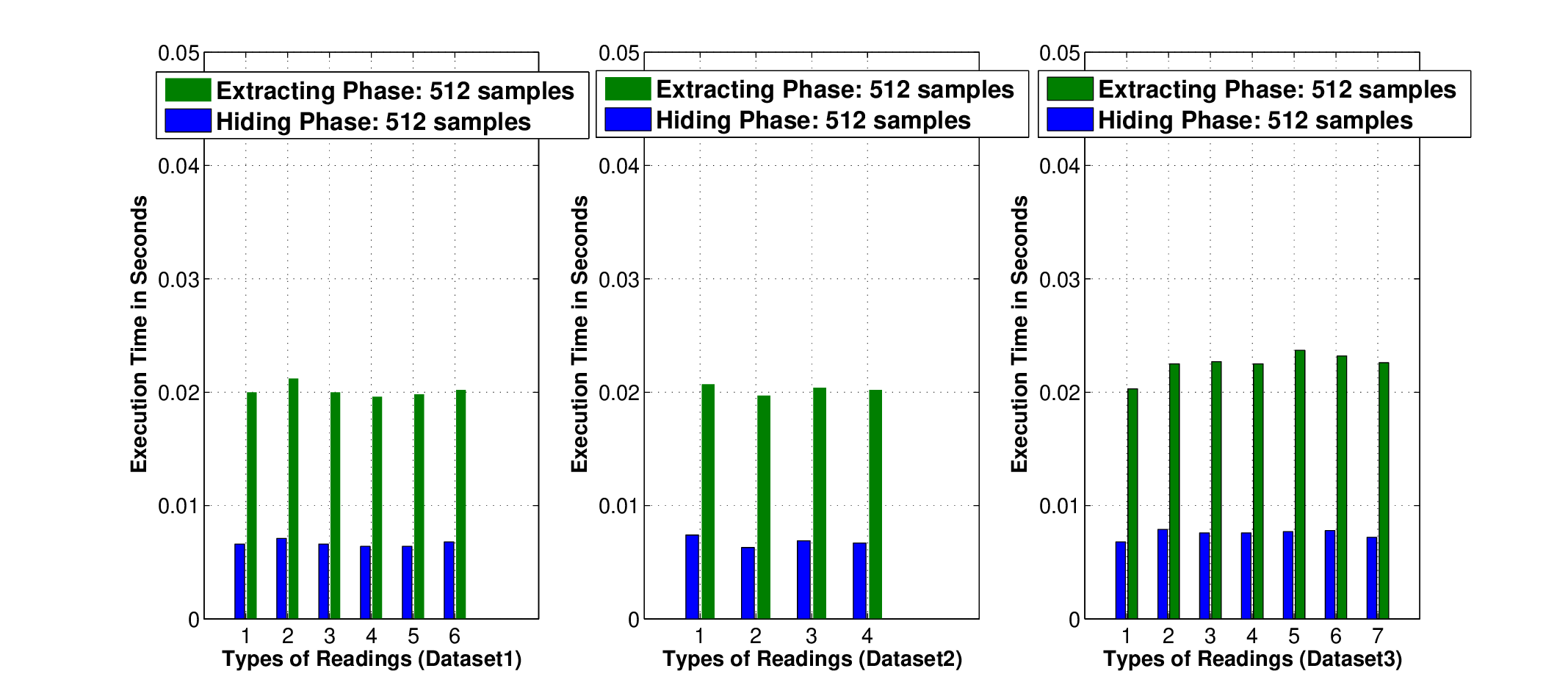}}
\caption{The required time and space needed by our algorithm to accomplish both hiding and retrieval process (a) 6 types of readings from the chemical dataset (b) 4 types of readings from the environmental dataset and (c) 7 examples of readings from the smart homes dataset.}
\label{fig:ExecutionTime}
\hfil
\end{figure*}
\subsection{IoT Resource Constraints }
Because of the IoT device limitations, the worst computational complexity (e.g. exponential) has been avoided during the developing of the algorithm's functionalities. Therefore, the worst-case complexity has been examined in two ways. (1) As shown in Table \ref{tb:bigOnotations},  big $\mathcal{O}$ notations have been used.  The majority of the functions are designed as linear tasks. It should be noted that Scattering and Sequence generation has been improved from quadratic to linear due to the utilization of radix sorting. From that, our algorithm is fully compatible with the new wireless network standards called IEEE 802.15.4/ZigBee \cite{zigbeestand:zheng2006zigbee}.  IEEE 802.15.4/ZigBee standards are already running algorithms with the computational complexity of logarithmic $\mathcal{O} (n\ log\ n)$  and quadratic $\mathcal{O} (n^2)$  \cite{zigbeesec:al2011aes,zigbeesec:bin2009}. (2) All streams in the repositories have been used and the execution time proved to be very low - $<$ 0.025 seconds - as presented in Fig \ref{fig:ExecutionTime}. 

The main tasks of the concealing model (See Table \ref{tb:bigOnotations}) that has to be executed by the remote IoTs are DCT and its inverse, random sequence formulation, private information scattering and embedding. Firstly, let's suppose $f(n)$ is $\mathcal{O}(g(n))$ if $f$ develops at farthest as $g$. Consequently, $f(n)$ = $\mathcal{O}(g(n))$ only if there occurs $c$, $n_0$ $\epsilon$ $\mathbb{R^+}$ such that for all $n \ge n_0$, $f(n) \le c.g(n)$. From that, for each 1D vector of a stream of length $n$ (i.e. 500 to 2048 in our experiments), the extreme complexity will be as follows. (i) $\mathcal{O}(n)$ for the time and space needed for DCT and its inverse \cite{dct:feig1992fast}. (ii) $\mathcal{O}(nk)$ for creating random sequences using a constant key  size $k$ $\epsilon$ $\mathbb{Z}$  after its enhancement to the linear problem. The space needed is the summation  of $n \quad and \quad k$.  (iii) The extreme required complexity for scattering the private information of length $m$ where at the lowest $< n/2$  is  $\mathcal{O}(m)$. (iv) $\mathcal{O}(nb)$ for concealing  $b$ $\epsilon$ $\mathbb{Z}$ (i.e. 1 to 10) bits in each coefficient of maximum length $n$, but the needed space is the accumulation  of both $n \quad and \quad m$.

It is obvious from the table that there is stability in the complexity in all three cases. This is due to the careful usage of the DCT in its simplistic form (i.e. 1D) and improving it by linear mathematical operations to achieve the required level of protection.

\begin{table}[!h]  
   \caption{ Algorithm Functionalities Computational Complexity}
   \label{tb:bigOnotations}
     \centerline{\begin{tabular}{| c | c | c | c | c |}
       \hline  
        Complexity &    \multicolumn{3}{|c|}{Time }  &  \multicolumn{1}{|c|}{Space }  \\ \hline 
       \pbox{30cm}{}  &\pbox{20cm}{Best}  & \pbox{20cm}{Average}  & \pbox{20cm}{Worst}  & \pbox{10cm}{Worst} \\ \hline  				
        DCT/Inverse & $\mathcal{O}(n)$  &  $\mathcal{O}(n)$ &  $\mathcal{O}(n)$ &   $\mathcal{O}(n)$\\ \hline
       Random Order & $\mathcal{O}(nk)$  &  $\mathcal{O}(nk)$  &   $\mathcal{O}(nk)$  &   $\mathcal{O}(n+k)$  \\ \hline
 Scramble Sensitive & $\mathcal{O}(m)$  &  $\mathcal{O}(m)$ &   $\mathcal{O}(m)$ &   $\mathcal{O}(m)$ \\ \hline
          Embedding & $\mathcal{O}(nb)$  &  $\mathcal{O}(nb)$ &   $\mathcal{O}(nb)$ &   $\mathcal{O}(n+m)$ \\ \hline              
     \end{tabular}}  
    \end{table} 

\section{Experiments and Results}\label{sec:experiments}
\subsection{Datasets}\label{subsec:dadaset}
To test the effectiveness of our model with different types of non-stationary streams data, three datasets have been examined. (1) Chemical dataset gathered and distributed by the University of California, Irvine Research Group \cite{ds:vergara2012chemical}. It contains extensive periodical readings over three years for six different volatile organic compounds: ethanol, ethylene, ammonia, acetaldehyde, acetone and toluene. (2) Environment dataset collected and published by Intel Berkeley Research Lab \cite{ds:Intel2004}. It offers detailed continues readings (i.e. every 31 seconds) over three months for four environmental characteristics: humidity, temperature, voltage and light. (3) Smart Homes dataset collected and published as a part of "Project Smart*" by Laboratory for Advanced System Software \cite{ds:barker2012smart,ds:smartproject2012}. It contains comprehensive periodical readings over three months (every minute) for different homes. The types of streams are power consumption, heat-index, inside/outside weather, inside/outside moisture and wind-cold. It also offers the utility consumption (electricity) from nearly 400 anonymous houses every minute for $(24 \times 30 \times 3) $ hours.

\begin{figure*}[!t]
\centerline
{\includegraphics[scale=0.5]{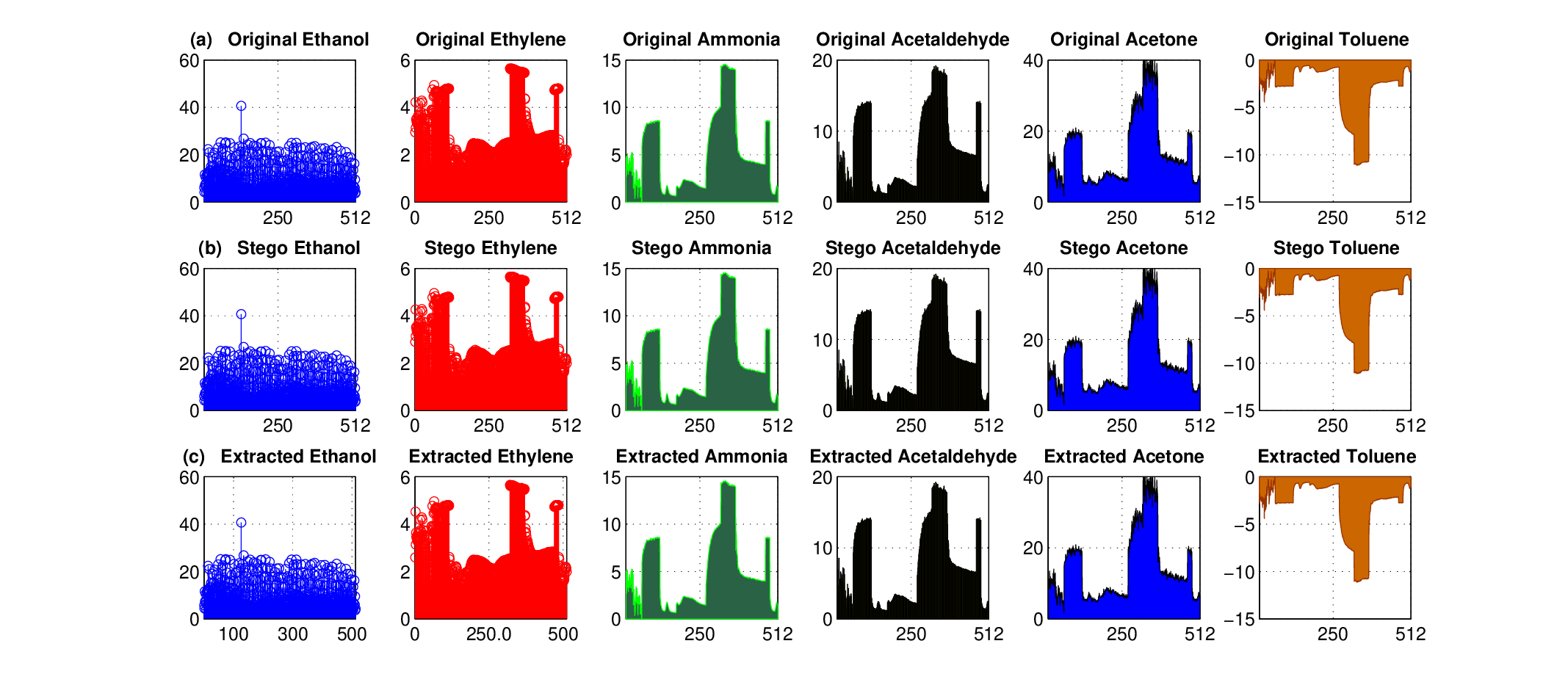}}
\caption{6 examples of chemical IoT streams: (a) Direct plot for original form (b) Stego form that contains the hidden private information (i.e. IDs) and (c) Recovered form (i.e. after removing the sensitive information).}
\label{fig:chemicaldataset}
\hfil
\end{figure*}
 \subsection{Experiments}
In this paper, all the above types of streams were used to thoroughly prove the feasibility of implementing the proposed algorithm on various collected streams. For all, experiments were done to conceal and recover the private data based on our algorithm steps presented in Sections \ref{subsec:hiding} and \ref{subsec:retrieve}. The secret bits was a set of values that have to be private such as location ID, geometric location, location picture, date and time  which all transformed into bits to be concealed inside the transmitted streams.
 
Our tests can be distributed into the following. (1) Hiding, which is done by remote IoT sensors to hide the remote locations' private data in their collected streams as presented in Section \ref{subsec:hiding}. (2) Private information recovery that is at the receiver's end as described in  \ref{subsec:retrieve}. Consequently, if the transmitted stego streams that contain the hidden private bits are sniffed or brutally altered  by unauthorized parties, (1) it will not disclose any private information and (2) it can be easily examined and verified.

\begin{figure*}[!t]
\centerline
{\includegraphics[scale=0.5]{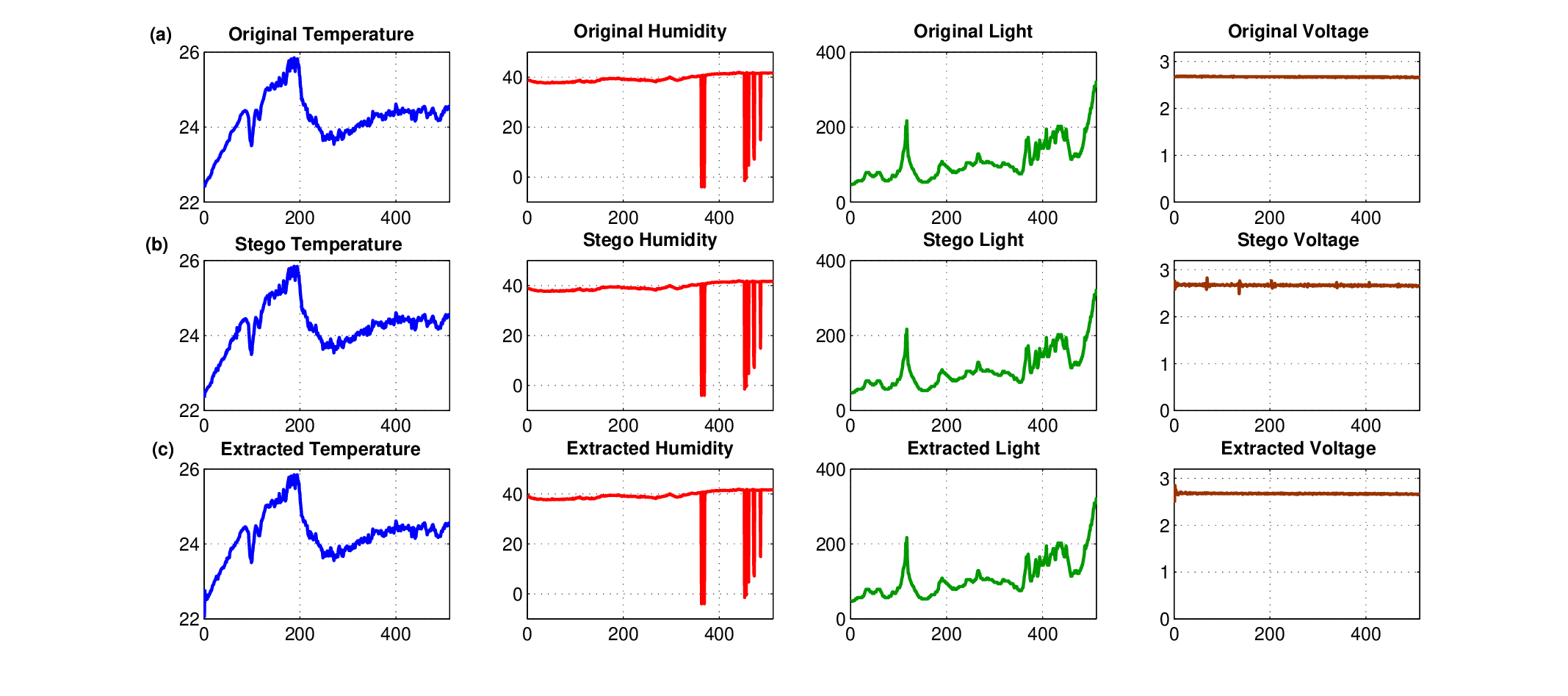}}

\caption{4 examples of environment IoT streams: (a) Direct plot for original form (b) Stego form that contains the hidden private information and (c) Extracted form (i.e. after removing the private information).}
\label{fig:Inteldataset}
\hfil
\end{figure*}

\begin{figure*}[!t]
\centerline
{\includegraphics[scale=0.5]{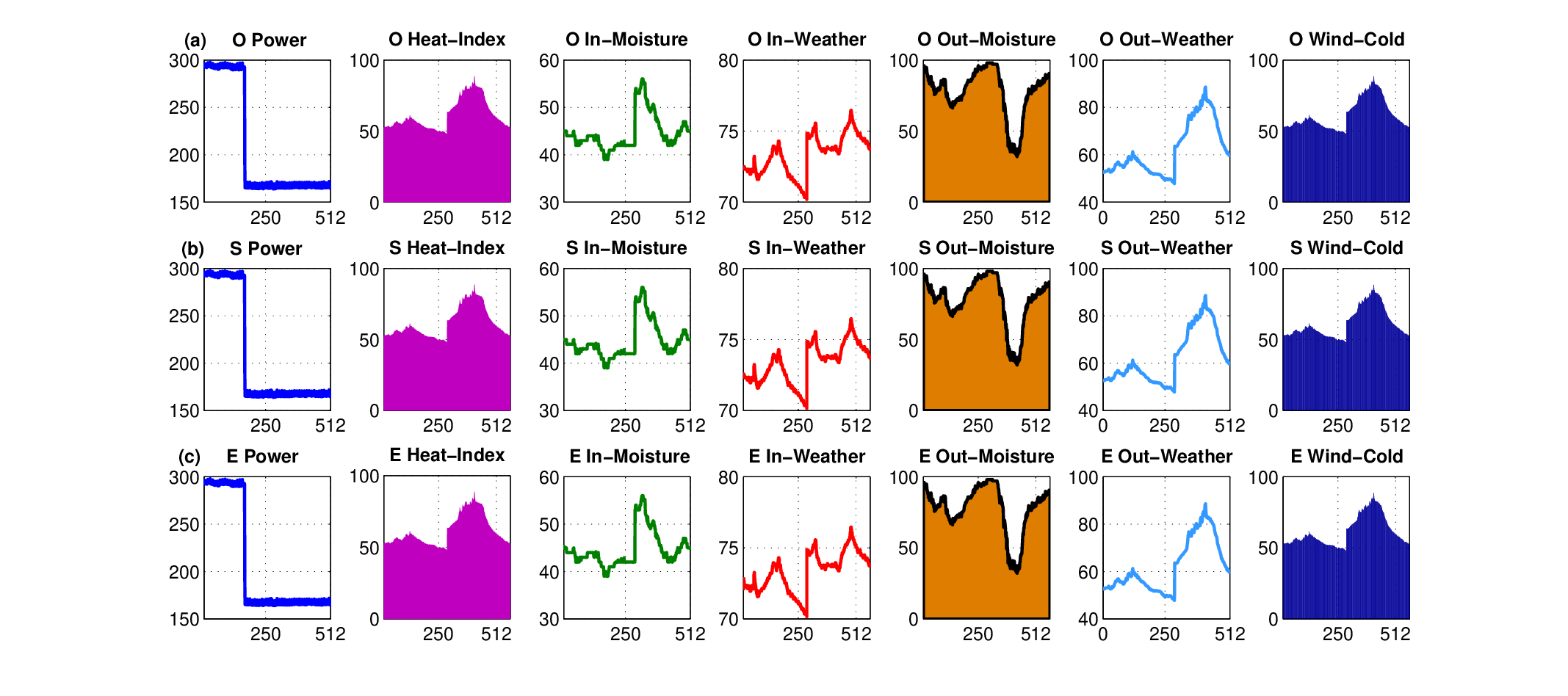}}
\caption{7 examples of smart homes IoT streams: (a) Direct plot for original form (b) stego form that contains the hidden private information (i.e. geometric location) and (c) Extracted form (i.e. after removing the private information).}
\label{fig:smartsataset}
\hfil
\end{figure*}

To obtain neutral and unbiased results, detailed experiments were performed with ranges of key sizes in addition to various collected streams having lengths (e.g. 512-to-4096). To highlight the extreme distortion impact, all detailed low sequence DCT values (i.e. around 95\% of the total coefficients) have been employed. For brevity, only a few examples of our results are presented, (1) Fig. \ref{fig:chemicaldataset} shows an instance using the chemical dataset of a plot of 6 ordinary streams  used to conceal private data, the stego and the extracted forms. (2) Fig. \ref{fig:Inteldataset}  shows an example using environment dataset of another plot of 4 ordinary streams, and the stego form before and after the extraction process. (3) Fig. \ref{fig:smartsataset} presents another example using smart homes dataset of a plot of 7 ordinary streams in addition to their stego and extracted forms. (4) Tables \ref{tb:prdchemicalDS}, \ref{tb:prdIntelDS} and \ref{tb:prdsmartDS} show the PRD measurements from all aforementioned datasets' streams between the actual and stego forms in addition to actual and the recovered forms.

\subsection{Discussion}
Despite the various sample lengths of the streams and various values' ranges, all PRDs are $<$ 1\%. This proves that the influence will only be to the less significant decimal digits (i.e. third or fourth) that typically are ignored. This guarantees that our technique will have stable and little distortion impact on the actually transmitted streams. On the contrary, it offers a promising solution with a paradigm shift for protecting the privacy of the transmitted private information as well as the originality of the periodically collected streams. The merits of this solution are as previously stated. (1) There are robust end-to-end privacy protection and authenticity where the hidden secured information can only be recovered and verified by legitimate recipients (e.g. operation centers), whereas others can only see the stego form which is almost similar to the original streams. (2) There is no increase in the actually transmitted streams. (3) There is no change to the original stream's form which helps the legitimate receivers to directly exploit operational administration such as cloud providers' services without disclosing private information. In other words, all mathematical operations can be directly applied to the transmitted stego form of streams even at intermediate hopes and cloud while maintaining the privacy and  authenticity.

\begin{table}[!h]  
   \caption{ PRD results for Chemical Data Set readings}
   \label{tb:prdchemicalDS}
     \centerline{\begin{tabular}{| c | c | c | c | c |}
       \hline  
        &    \multicolumn{2}{|c|}{Batch 9}  &  \multicolumn{2}{|c|}{Batch 10}  \\ \hline 
       \pbox{20cm}{Segment\\ No}  &\pbox{20cm}{PRD \% \\ Stego}  & \pbox{20cm}{PRD \%\\ Recovered}  & \pbox{20cm}{PRD \% \\ Stego}  & \pbox{20cm}{PRD \%\\ Recovered} \\ \hline  				
         1 & 0.1989  &  0.2437 &   0.1887 &   0.2284\\ \hline
         2 & 0.3068  &  0.3611 &   0.6665 &   0.7834 \\ \hline
         3 & 0.0520  &  0.0638 &   0.3157 &   0.4126 \\ \hline
         4 & 0.0399  &  0.0477 &   0.2304 &   0.2858 \\ \hline
         5 & 0.0305  &  0.0377 &   0.1206 &   0.1498 \\ \hline
         6 & 0.6483  &  0.6485 &   0.4974 &   0.5928\\ \hline
         7 & 0.0432  &  0.0531 &   0.1839 &   0.2159\\ \hline
         8 & 0.0937  &  0.1157 &   0.1042 &   0.1266 \\ \hline
         9 & 0.3214  &  0.3979 &   0.6452 &   0.8083 \\ \hline
        10 & 0.1954  &  0.2220 &   0.2375 &   0.2842 \\ \hline
        11 & 0.1887  &  0.2284 &   0.2376 &   0.2789 \\ \hline
        12 & 0.5818  &  0.6941 &   0.9745 &   0.2284\\ \hline
        13 & 0.3643  &  0.4494 &   0.3766 &   0.4524 \\ \hline
        14 & 0.4365  &  0.5152 &   0.4162 &   0.4973 \\ \hline
        15 & 0.3679  &  0.4631 &   0.6326 &   0.7817 \\ \hline
        16 & 0.1453  &  0.1797 &   0.1804 &   0.2268\\ \hline
     \end{tabular}}  
    \end{table}     
    
    \begin{table}[!h]  
       \caption{ PRD results for Intel Data Set readings}
       \label{tb:prdIntelDS}
         \centerline{\begin{tabular}{| c | c | c | c | c |}
           \hline 
           &    \multicolumn{2}{|c|}{Temperature}  &  \multicolumn{2}{|c|}{Humidity}  \\ \hline 
           \pbox{20cm}{Segment\\ No}  &\pbox{20cm}{PRD \% \\ Stego}  & \pbox{20cm}{PRD \%\\ Recovered}  & \pbox{20cm}{PRD \% \\ Stego}  & \pbox{20cm}{PRD \%\\ Recovered} \\ \hline 				
             1 & 0.0708  &  0.1033 &   0.3492 &   0.3514\\ \hline
             2 & 0.0723  &  0.1030 &   0.0538 &   0.0674 \\ \hline
             3 & 0.0711  &  0.0878 &   0.0514 &   0.0650 \\ \hline
             4 & 0.0707  &  0.0906 &   0.0734 &   0.0889 \\ \hline
             5 & 0.0779  &  0.1081 &   0.2528 &   0.2255 \\ \hline
             6 & 0.0787  &  0.1086 &   0.0549 &   0.0662\\ \hline
             7 & 0.0700  &  0.0964 &   0.1544 &   0.1572\\ \hline
             8 & 0.0686  &  0.0963 &   0.0573 &   0.0654 \\ \hline
             &    \multicolumn{2}{|c|}{Light}  &  \multicolumn{2}{|c|}{Voltage}  \\ \hline 
                        \pbox{20cm}{Segment\\ No}  &\pbox{20cm}{PRD \% \\ Stego}  & \pbox{20cm}{PRD \%\\ Recovered}  & \pbox{20cm}{PRD \% \\ Stego}  & \pbox{20cm}{PRD \%\\ Recovered} \\ \hline 
             9 & 0.0220  &  0.0253 &   0.0360 &   0.0611 \\ \hline
            10 & 0.0606  &  0.0750 &   0.6936 &   0.4837 \\ \hline
            11 & 0.0331  &  0.0376 &   0.7041 &   0.4613\\ \hline
            12 & 0.4276  &  0.4276 &   0.7264 &   0.4886\\ \hline
            13 & 0.0138  &  0.0170 &   0.8326 &   0.9902 \\ \hline
            14 & 0.0521  &  0.0645 &   0.7328 &   0.5115 \\ \hline
            15 & 0.0511  &  0.0631 &   0.7130 &   0.4661\\ \hline
            16 & 0.0182  &  0.0218 &   0.7053 &   0.4382\\ \hline
         \end{tabular}} 
        \end{table} 
    
     \begin{table}[!h]  
           \caption{ PRD results for Smart Project Data Set readings}
           \label{tb:prdsmartDS}
             \centerline{\begin{tabular}{| c | c | c | c | c |}
               \hline 
               &    \multicolumn{2}{|c|}{Power}  &  \multicolumn{2}{|c|}{Heat-Index}  \\ \hline 
               \pbox{20cm}{Segment\\ No}  &\pbox{20cm}{PRD \% \\ Stego}  & \pbox{20cm}{PRD \%\\ Recovered}  & \pbox{20cm}{PRD \% \\ Stego}  & \pbox{20cm}{PRD \%\\ Recovered} \\ \hline 				
                 1 & 0.1019  &  0.1019 &   0.0346 &   0.0411\\ \hline
                 2 & 0.0076  &  0.0091 &   0.0334 &   0.0409 \\ \hline
                 3 & 0.0077  &  0.0094 &   0.0337 &   0.0399 \\ \hline
                 4 & 0.0132  &  0.0155 &   0.0311 &   0.0391 \\ \hline
                 5 & 0.0131  &  0.0158 &   0.9506 &   0.9509 \\ \hline
                 6 & 0.7173  &  0.7173 &   0.0321 &   0.0388\\ \hline
                 7 & 0.1148  &  0.1148 &   0.0337 &   0.0416\\ \hline
                 8 & 0.0901  &  0.0901 &   0.0324 &   0.0377 \\ \hline
                 &    \multicolumn{2}{|c|}{Wind-Cold}  &  \multicolumn{2}{|c|}{In-Weather}  \\ \hline 
                            \pbox{20cm}{Segment\\ No}  &\pbox{20cm}{PRD \% \\ Stego}  & \pbox{20cm}{PRD \%\\ Recovered}  & \pbox{20cm}{PRD \% \\ Stego}  & \pbox{20cm}{PRD \%\\ Recovered} \\ \hline 
                 9 & 0.0343  &  0.0412 &   0.0292 &   0.0362 \\ \hline
                10 & 0.0346  &  0.0389 &   0.0293 &   0.0374 \\ \hline
                11 & 0.0335  &  0.0408 &   0.0296 &   0.0369\\ \hline
                12 & 0.0322  &  0.0376 &   0.0286 &   0.0361\\ \hline
                13 & 0.0314  &  0.0376 &   0.0294 &   0.0364 \\ \hline
                14 & 0.0310  &  0.0398 &   0.0292 &   0.0349 \\ \hline
                15 & 0.0341  &  0.0411 &   0.0301 &   0.0368\\ \hline
                16 & 0.0309  &  0.0369 &   0.0297 &   0.0362\\ \hline
 &    \multicolumn{2}{|c|}{In-Moisture}  &  \multicolumn{2}{|c|}{Out-Moisture}  \\ \hline 
                            \pbox{20cm}{Segment\\ No}  &\pbox{20cm}{PRD \% \\ Stego}  & \pbox{20cm}{PRD \%\\ Recovered}  & \pbox{20cm}{PRD \% \\ Stego}  & \pbox{20cm}{PRD \%\\ Recovered} \\ \hline 
                17 & 0.0506  &  0.0609 &   0.0298 &   0.0356 \\ \hline
                18 & 0.0507  &  0.0626 &   0.0306 &   0.0357 \\ \hline
                19 & 0.0479  &  0.0600 &   0.0303 &   0.0362\\ \hline
                20 & 0.0524  &  0.0615 &   0.0317 &   0.0407\\ \hline
                21 & 0.0510  &  0.0596 &   0.0296 &   0.0364 \\ \hline
                22 & 0.0459  &  0.0577 &   0.0304 &   0.0366 \\ \hline
                23 & 0.0507  &  0.0592 &   0.0268 &   0.0320\\ \hline
                24 & 0.0434  &  0.0534 &   0.0267 &   0.0336\\ \hline
             \end{tabular}} 
            \end{table} 

\begin{table}[!h]  
	\caption{ Comparison of IoT streams stego-based solutions and our approach}
	\label{tb:comparision}
	\begin{tabular}{| c | c | c | c |}
			\hline  
			
			Feature  &  \cite{stego:abuadbba2015robust}&  \cite{stego:abuadbba2016resilient} & Our approach\\ \hline  				
			Signal processing & Walsh-Hadamard  &  Wavelet &  DCT \\ \hline
			Complexity & $\mathcal{O} (n\ log\ n)$  &  $\mathcal{O} (n)^2$  &  $\mathcal{O} (n)$   \\ \hline
			Hiding per/coefficient & 5  & 6  &   10  \\ \hline            
		\end{tabular}  
	\end{table} 
\subsection{Comparison with Existing Models}
The comparison focuses on two folds. Firstly, the superiority of this model over our previous steganographic-based techniques in the context of IoT streams \cite{stego:abuadbba2015wavelet,stego:abuadbba2015robust,stego:abuadbba2016resilient} which has been summarized in Table \ref{tb:comparision}. Secondly, there are recent works proposed by  Vongurai and Phimoltares \cite{stego:jpegimg2012:Vongurai}, Biswas et. al. \cite{stego:colouredimg2013:Biswas}, and Bhaskar et. al. \cite{stego:video2013:Bhaskar} that have similar signal processing (i.e. DCT) with a different context (i.e. multimedia domain). Therefore, our work is compared with these three recent techniques where (1) in \cite{stego:jpegimg2012:Vongurai}, the authors used DCT decomposition to conceal a secret message inside transmitted JPEG image; (2) the authors in \cite{stego:colouredimg2013:Biswas} utilized DCT to hide a secret data inside a colored image using a predefined password; (3) the authors in \cite{stego:video2013:Bhaskar} applied DCT to hide a secret content inside transmitted MPEG-4 videos. 

\begin{figure*}[!t]
 \centerline
 {\includegraphics[scale=0.4]{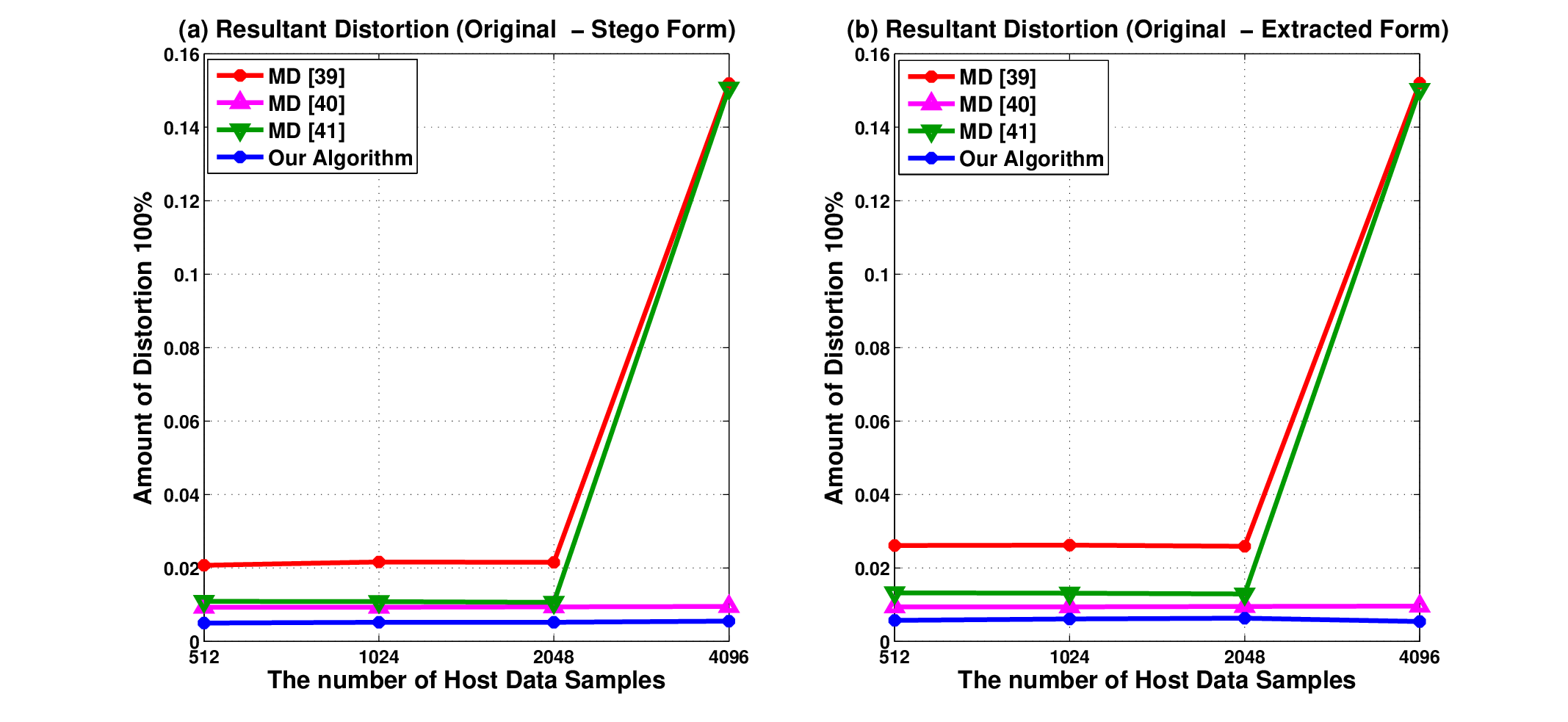}}
 \caption{ Resultant distortion after applying our algorithm and the algorithms in \cite{stego:jpegimg2012:Vongurai,stego:colouredimg2013:Biswas,stego:video2013:Bhaskar} on the same amount of numerical data. (a) PRDs between the original and stego form, (b) PRDs between the original and the extracted form. }
 \label{fig:Distortion_comparison}
 \hfil
 \end{figure*}

\begin{figure}[!h]
 \centerline
 {\includegraphics[scale=0.25]{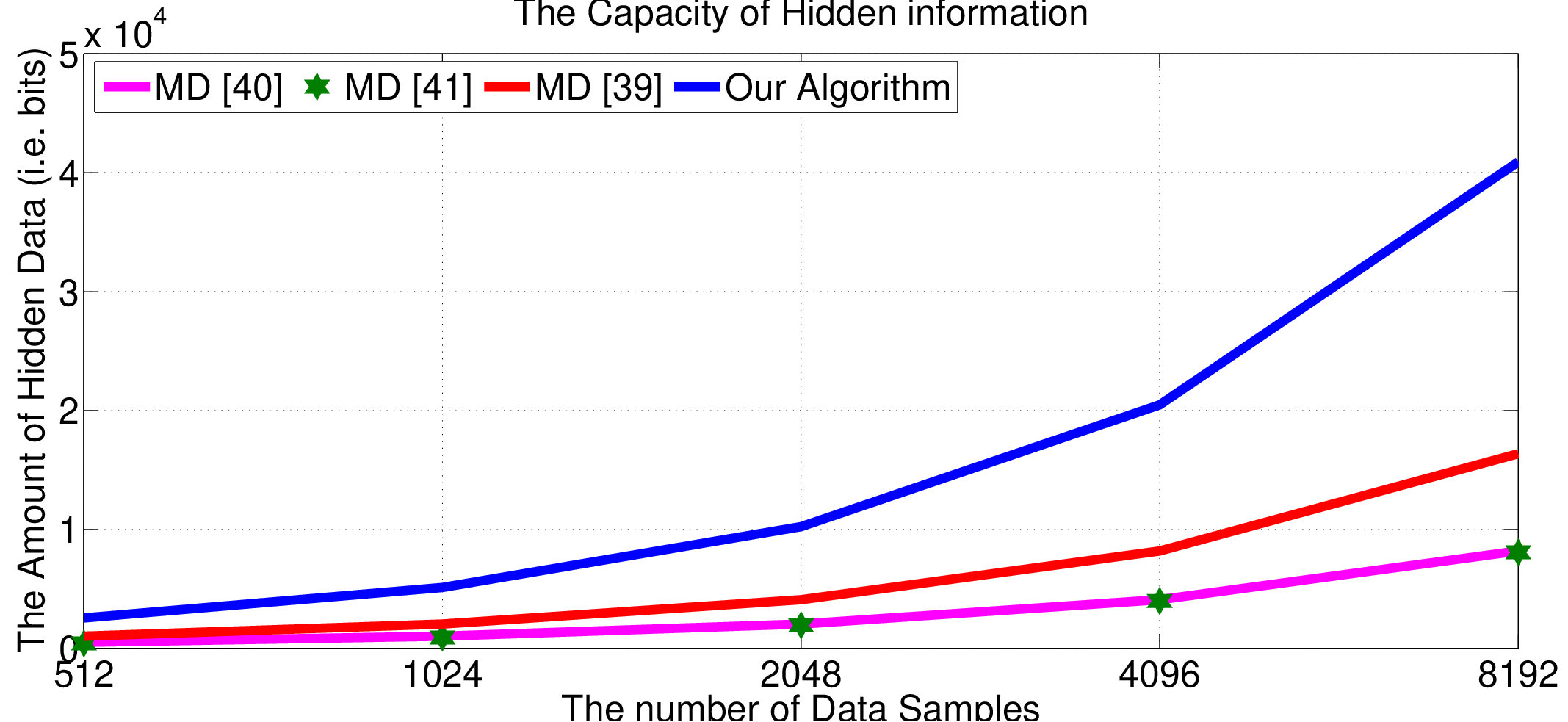}}
 \caption{ Comparison  for the maximum amount of secret data that can be hidden between our algorithm  and the algorithms in \cite{stego:jpegimg2012:Vongurai,stego:colouredimg2013:Biswas,stego:video2013:Bhaskar}.}
 \label{fig:CapacityofHiddenData}
 
 \end{figure}

The proposed technique has the following improvements. 
\begin{enumerate}
\item After experimenting with the same number of the transmitted collected streams,  it is clear from Fig. \ref{fig:CapacityofHiddenData} that the capacity of the hidden secret information is much higher in our algorithm than in models \cite{stego:jpegimg2012:Vongurai,stego:colouredimg2013:Biswas,stego:video2013:Bhaskar} where up to 10 bits can be concealed in each DCT value because of exploiting  the least significant DCT coefficients, whereas just 1 to 2 bits can be concealed in their algorithms.
\item From Fig.  \ref{fig:Distortion_comparison}, it should be noticed that  our algorithm has less resultant distortion than other algorithms because it has been designed specifically to be aware of the sensitivity of the important features of the numeric data, whereas other algorithms designed to  wider content (e.g. images and videos).
\item Most significantly, our model is strongly secure compared to  the models in \cite{stego:jpegimg2012:Vongurai,stego:colouredimg2013:Biswas,stego:video2013:Bhaskar} due to their static and immediate secret bits distribution  in the absence of a strong key or using just a simple password, but in our model various security layers are implemented which are scattering the resultant DCT values, obfuscating the private bits and producing a random sequence derived from the key to dynamically distribute the bits among arbitrary  coefficients.
\end{enumerate}

 
\section{Conclusion}
In this work, a new secure steganographic technique has been introduced to protect privately transmitted information with streams by distributing them randomly employing a secret key. This will provide (1) robust end-to-end privacy protection for sensitive information, and (2) strong proof of originality for the normal streams. To ensure the highest hiding capacity, DCT is applied to compose the readings into a group of coefficients. To guarantee minimum distortion, only the least significant values are used. To strengthen the security, a key is used to (1) only obfuscate the private information, (2) reorganize the coefficients into a $2D$  $M \times N$, and (3) create a random order in the form of $2D$  $\widetilde M \times \widetilde N$ used in the hiding process. The resultant distortion has been carefully measured in all stages - the original, the stego, and the extracted forms - using a well-known measurement called PRD. After thorough experimentation on three different types of streams (i.e. chemical, environment and smart homes) it has been proven that our model has little impact on the actual readings ($<$ 1\%). Also, computational complexity has been proven to be much lighter $\mathcal{O} (n)$ compared to existing work. 

\bibliographystyle{unsrt}
\bibliography{References}

\begin{thebibliography}{10}

\bibitem{int:yick2008wireless}
Jennifer Yick, Biswanath Mukherjee, and Dipak Ghosal.
\newblock Wireless sensor network survey.
\newblock {\em Computer networks}, 52(12):2292--2330, 2008.

\bibitem{ibaida2021privacy}
Ayman Ibaida, A.~Abuadbba, and Naveen Chilamkurti.
\newblock Privacy-preserving compression model for efficient iomt ecg sharing.
\newblock {\em Computer Communications}, 166:1--8, 2021.

\bibitem{baniata2021energy}
Mohammad Baniata, Haftu~Tasew Reda, Naveen Chilamkurti, and A.~Abuadbba.
\newblock Energy-efficient hybrid routing protocol for iot communication
  systems in 5g and beyond.
\newblock {\em Sensors}, 21(2):537, 2021.

\bibitem{PrivToStatic:kargupta2003privacy}
Hillol Kargupta, Souptik Datta, Qi~Wang, and Krishnamoorthy Sivakumar.
\newblock On the privacy preserving properties of random data perturbation
  techniques.
\newblock In {\em Data Mining, 2003. ICDM 2003. Third IEEE International
  Conference on}, pages 99--106. IEEE, 2003.

\bibitem{PrivToStatic:liu2006random}
Kun Liu, Hillol Kargupta, and Jessica Ryan.
\newblock Random projection-based multiplicative data perturbation for privacy
  preserving distributed data mining.
\newblock {\em Knowledge and Data Engineering, IEEE Transactions on},
  18(1):92--106, 2006.

\bibitem{PrivToStatic:machanavajjhala2007diversity}
Ashwin Machanavajjhala, Daniel Kifer, Johannes Gehrke, and Muthuramakrishnan
  Venkitasubramaniam.
\newblock l-diversity: Privacy beyond k-anonymity.
\newblock {\em ACM Transactions on Knowledge Discovery from Data (TKDD)},
  1(1):3, 2007.

\bibitem{PrivStream:li2007hiding}
Feifei Li, Jimeng Sun, Spiros Papadimitriou, George~A Mihaila, and Ioana
  Stanoi.
\newblock Hiding in the crowd: Privacy preservation on evolving streams through
  correlation tracking.
\newblock In {\em ICDE}, volume~1, page~2, 2007.

\bibitem{sensorissue:li2009privacy}
Na~Li, Nan Zhang, Sajal~K Das, and Bhavani Thuraisingham.
\newblock Privacy preservation in wireless sensor networks: A state-of-the-art
  survey.
\newblock {\em Ad Hoc Networks}, 7(8):1501--1514, 2009.

\bibitem{tradcrypto:yang2008sdap}
Yi~Yang, Xinran Wang, Sencun Zhu, and Guohong Cao.
\newblock Sdap: A secure hop-by-hop data aggregation protocol for sensor
  networks.
\newblock {\em ACM Transactions on Information and System Security (TISSEC)},
  11(4):18, 2008.

\bibitem{tradcrypto:lu2013lightweight}
Rongxing Lu, Xiaodong Lin, Zhiguo Shi, and Xuemin~(Sherman) Shen.
\newblock A lightweight conditional privacy-preservation protocol for vehicular
  traffic-monitoring systems.
\newblock {\em IEEE Intelligent Systems}, 28(3):62--65, 2013.

\bibitem{tradcrypto2013:Junzuo}
Junzuo Lai, R.H. Deng, Chaowen Guan, and Jian Weng.
\newblock Attribute-based encryption with verifiable outsourced decryption.
\newblock {\em Information Forensics and Security, IEEE Transactions on},
  8(8):1343--1354, Aug 2013.

\bibitem{tradcrypto:puttaswamy2014preserving}
Krishna~PN Puttaswamy, Shiyuan Wang, Troy Steinbauer, Divyakant Agrawal, Amr
  El~Abbadi, Christopher Kruegel, and Ben~Y Zhao.
\newblock Preserving location privacy in geosocial applications.
\newblock {\em Mobile Computing, IEEE Transactions on}, 13(1):159--173, 2014.

\bibitem{tradcrypto2014:Han}
J.~Han, W.~Susilo, Y.~Mu, J.~Zhou, and M.H.A. Au.
\newblock Improving privacy and security in decentralized ciphertext-policy
  attribute-based encryption.
\newblock {\em Information Forensics and Security, IEEE Transactions on},
  10(3):665--678, March 2015.

\bibitem{tradcrypto2017:song}
T.~{Song}, R.~{Li}, B.~{Mei}, J.~{Yu}, X.~{Xing}, and X.~{Cheng}.
\newblock A privacy preserving communication protocol for iot applications in
  smart homes.
\newblock {\em IEEE Internet of Things Journal}, 4(6):1844--1852, Dec 2017.

\bibitem{hom:liang2013udp}
Xiaohui Liang, Xu~Li, Rongxing Lu, Xiaodong Lin, and Xuemin Shen.
\newblock Udp: Usage-based dynamic pricing with privacy preservation for smart
  grid.
\newblock {\em Smart Grid, IEEE Transactions on}, 4(1):141--150, 2013.

\bibitem{hom:lien2013novel}
I-Ting Lien, Yu-Hsun Lin, Jyh-Ren Shieh, and Ja-Ling Wu.
\newblock A novel privacy preserving location-based service protocol with
  secret circular shift for k-nn search.
\newblock {\em Information Forensics and Security, IEEE Transactions on},
  8(6):863--873, 2013.

\bibitem{hom:saleem2014aggregation}
Noman Saleem, Saed Alrabaee, Fawaz~A Khasawneh, and Mahmoud Khasawneh.
\newblock Aggregation function using homomorphic encryption in participating
  sensing application.
\newblock In {\em Computer Science and Information Technology (CSIT), 2014 6th
  International Conference on}, pages 166--171. IEEE, 2014.

\bibitem{hom:kumar2015secure}
Manish Kumar, Shekhar Verma, and Kusum Lata.
\newblock Secure data aggregation in wireless sensor networks using homomorphic
  encryption.
\newblock {\em International Journal of Electronics}, 102(4):690--702, 2015.

\bibitem{homocrypto2019:Bor}
C.~{Borrego}, M.~{Amadeo}, A.~{Molinaro}, and R.~H. {Jhaveri}.
\newblock Privacy-preserving forwarding using homomorphic encryption for
  information-centric wireless ad hoc networks.
\newblock {\em IEEE Communications Letters}, 23(10):1708--1711, Oct 2019.

\bibitem{wat:zhang2008secure}
Wei Zhang, Yonghe Liu, Sajal~K Das, and Pradip De.
\newblock Secure data aggregation in wireless sensor networks: a watermark
  based authentication supportive approach.
\newblock {\em Pervasive and Mobile Computing}, 4(5):658--680, 2008.

\bibitem{randper:kargupta2003privacy}
Hillol Kargupta, Souptik Datta, Qi~Wang, and Krishnamoorthy Sivakumar.
\newblock On the privacy preserving properties of random data perturbation
  techniques.
\newblock In {\em Data Mining, 2003. ICDM 2003. Third IEEE International
  Conference on}, pages 99--106. IEEE, 2003.

\bibitem{randper:huang2005deriving}
Zhengli Huang, Wenliang Du, and Biao Chen.
\newblock Deriving private information from randomized data.
\newblock In {\em Proceedings of the 2005 ACM SIGMOD international conference
  on Management of data}, pages 37--48. ACM, 2005.

\bibitem{k-anonymity:gedik2005location}
Bugra Gedik and Ling Liu.
\newblock Location privacy in mobile systems: A personalized anonymization
  model.
\newblock In {\em Distributed Computing Systems, 2005. ICDCS 2005. Proceedings.
  25th IEEE International Conference on}, pages 620--629. IEEE, 2005.

\bibitem{homcrisis:groat2011kipda}
Michael~M Groat, Wenbo He, and Stephanie Forrest.
\newblock Kipda: k-indistinguishable privacy-preserving data aggregation in
  wireless sensor networks.
\newblock In {\em INFOCOM, 2011 Proceedings IEEE}, pages 2024--2032. IEEE,
  2011.

\bibitem{k-anonymity:chow2011privacy}
Chi-Yin Chow, Mohamed~F Mokbel, and Tian He.
\newblock A privacy-preserving location monitoring system for wireless sensor
  networks.
\newblock {\em Mobile Computing, IEEE Transactions on}, 10(1):94--107, 2011.

\bibitem{stego:provos2003hide}
Niels Provos and Peter Honeyman.
\newblock Hide and seek: An introduction to steganography.
\newblock {\em Security \& Privacy, IEEE}, 1(3):32--44, 2003.

\bibitem{stego:cox2002digital}
Ingemar~J Cox, Matthew~L Miller, Jeffrey~A Bloom, and Chris Honsinger.
\newblock {\em Digital watermarking}, volume~53.
\newblock Springer, 2002.

\bibitem{stego:abuadbba2015wavelet}
A.~Abuadbba and Ibrahim Khalil.
\newblock Wavelet based steganographic technique to protect household
  confidential information and seal the transmitted smart grid readings.
\newblock {\em Information Systems}, 53:224--236, 2015.

\bibitem{stego:abuadbba2015robust}
A.~Abuadbba, Ibrahim Khalil, and Mohammed Atiquzzaman.
\newblock Robust privacy preservation and authenticity of the collected data in
  cognitive radio network—walsh--hadamard based steganographic approach.
\newblock {\em Pervasive and Mobile Computing}, 22:58--70, 2015.

\bibitem{stego:abuadbba2016resilient}
A.~Abuadbba, Ibrahim Khalil, Ayman Ibaida, and Mohammed Atiquzzaman.
\newblock Resilient to shared spectrum noise scheme for protecting cognitive
  radio smart grid readings- bch based steganographic approach.
\newblock {\em Ad Hoc Networks}, 41:30--46, 2016.

\bibitem{k-anonymityCris:machanavajjhala2007diversity}
Ashwin Machanavajjhala, Daniel Kifer, Johannes Gehrke, and Muthuramakrishnan
  Venkitasubramaniam.
\newblock l-diversity: Privacy beyond k-anonymity.
\newblock {\em ACM Transactions on Knowledge Discovery from Data (TKDD)},
  1(1):3, 2007.

\bibitem{dct:feig1992fast}
Ephraim Feig and Shmuel Winograd.
\newblock Fast algorithms for the discrete cosine transform.
\newblock {\em Signal Processing, IEEE Transactions on}, 40(9):2174--2193,
  1992.

\bibitem{dct:rao1990discrete}
Kamisetty~Ramamohan Rao, Ping Yip, and K~Ramamohan Rao.
\newblock {\em Discrete cosine transform: algorithms, advantages,
  applications}, volume 226.
\newblock Academic press Boston, 1990.

\bibitem{dct:ahmed1974discrete}
Nasir Ahmed, T~Natarajan, and Kamisetty~R Rao.
\newblock Discrete cosine transform.
\newblock {\em Computers, IEEE Transactions on}, 100(1):90--93, 1974.

\bibitem{prd:cetin2000compression}
A~Enis Cetin and Hayrettin K{\"o}ymen.
\newblock Compression of digital biomedical signals.
\newblock {\em The Biomedical Engineering Handbook: Second Edition. Joseph D.
  Bonzino, Ed. CRC Press LLC}, 2000.

\bibitem{zigbeestand:zheng2006zigbee}
Li~Zheng.
\newblock Zigbee wireless sensor network in industrial applications.
\newblock In {\em SICE-ICASE International Joint Conference}, volume~10, page
  2006, 2006.

\bibitem{zigbeesec:al2011aes}
Saif Al-alak, Zuriati Ahmed, Azizol Abdullah, and Shamala Subramiam.
\newblock Aes and ecc mixed for zigbee wireless sensor security.
\newblock {\em computing}, 1:5, 2011.

\bibitem{zigbeesec:bin2009}
Bin Yang.
\newblock Study on security of wireless sensor network based on zigbee
  standard.
\newblock In {\em Computational Intelligence and Security, 2009. CIS '09.
  International Conference on}, volume~2, pages 426--430, Dec 2009.

\bibitem{ds:vergara2012chemical}
Alexander Vergara, Shankar Vembu, Tuba Ayhan, Margaret~A Ryan, Margie~L Homer,
  and Ram{\'o}n Huerta.
\newblock Chemical gas sensor drift compensation using classifier ensembles.
\newblock {\em Sensors and Actuators B: Chemical}, 166:320--329, 2012.

\bibitem{ds:Intel2004}
P~Bodik et~al.
\newblock Intel berkeley research lab.
\newblock \url{db.csail.mit.edu/labdata/labdata.html}, 2004.

\bibitem{ds:barker2012smart}
Sean Barker, Aditya Mishra, David Irwin, Emmanuel Cecchet, Prashant Shenoy, and
  Jeannie Albrecht.
\newblock Smart*: An open data set and tools for enabling research in
  sustainable homes.
\newblock {\em SustKDD, August}, 2012.

\bibitem{ds:smartproject2012}
Sean Barker, Aditya Mishra, David Irwin, Emmanuel Cecchet, Prashant Shenoy, and
  Jeannie Albrecht.
\newblock Smart project.
\newblock \url{http://traces.cs.umass.edu/index.php/Smart/Smart}, 2012.

\bibitem{stego:jpegimg2012:Vongurai}
N.~Vongurai and S.~Phimoltares.
\newblock Frequency-based steganography using 32x32 interpolated quantization
  table and discrete cosine transform.
\newblock In {\em Computational Intelligence, Modelling and Simulation
  (CIMSiM), 2012 Fourth International Conference on}, pages 249--253, Sept
  2012.

\bibitem{stego:colouredimg2013:Biswas}
R.~Biswas, S.~Mukherjee, and S.K. Bandyopadhyay.
\newblock Dct domain encryption in lsb steganography.
\newblock In {\em Computational Intelligence and Communication Networks (CICN),
  2013 5th International Conference on}, pages 405--408, Sept 2013.

\bibitem{stego:video2013:Bhaskar}
T.~Bhaskar and D.~Vasumathi.
\newblock A reversible data embedding scheme for mpeg-4 video using non-zero ac
  coefficients of dct.
\newblock In {\em Computational Intelligence and Computing Research (ICCIC),
  2013 IEEE International Conference on}, pages 1--4, Dec 2013.

\end{thebibliography}

\end{document}